\documentclass[12pt]{iopart}
\usepackage[latin1]{inputenc}
\usepackage{graphicx}
\usepackage{setstack}
\usepackage{subfigure}
\usepackage{iopams} 
\usepackage{dcolumn}
\newcolumntype{d}[1]{D{.}{.}{#1}}

\newcommand{\chem}[1]{{$\fontdimen16\tensy=3.0pt\fontdimen17\tensy=3.0pt \mathrm{#1}$}}
\def\la{{\langle\, }}
\def\ra{{\,\rangle }}

\def\infintd3r{ \int_{-\infty}^\infty d^3r\,}
\def\intd3r{ \int d^3r\,}

\def\laplace1d{\frac{d^2}{dx^2}}
\def\plaplace1d{\frac{d^2}{d{x'}^2}}

\def\padr2{\frac{\partial^2}{\partial r^2}}

\def\bei{\begin{itemize}}
\def\eei{\end{itemize}}
\def\bee{\begin{enumerate}}
\def\eee{\end{enumerate}}

\def\TDSE{time-dependent Schr\"odinger equation}

\def\octopus{OCTOPUS}
\newfont{\tensy}{cmsy10}
\newcommand{\im}[1]{\,\mathrm{Im}#1}
\newcommand{\re}[1]{\,\mathrm{Re}#1}

\def\bepsilon{\boldsymbol \epsilon}  
\def\bmu{\boldsymbol \mu} 
\def\balpha{\boldsymbol \alpha}
\newcommand{\aref}[1]{\ref{#1}}

\def\bea{\begin{eqnarray}}
\def\eea{\end{eqnarray}}
\def\ben{\begin{equation}}
\def\een{\end{equation}}

\def\sss{\scriptscriptstyle\rm}

\def\1var{(\bx_1...\bx\N)}

\def\half{\frac{1}{2}}

\def\brr{{\bf r}}

\def\bx{{\bf x}}

\def\N{_{\sss N}}

\def\td{time-dependent~}

\def\sph_int{ {\int d^3 r}}

\begin{document}
\title[Quantum Optimal Control Theory]{Quantum Optimal Control Theory}
\author{J. Werschnik and E.K.U. Gross}
\address{Institut für Theoretische Physik, Freie Universität  Berlin, Arnimallee 14, 14195 Berlin, Germany}
\date{\today}
\ead{jan.werschnik@physik.fu-berlin.de}
\begin{abstract}
The control of quantum dynamics via specially tailored laser pulses is a long-standing goal in physics and chemistry. Partly, this dream has come true, as sophisticated pulse shaping experiments allow to coherently control product ratios of chemical reactions. The theoretical design of the laser pulse to transfer an initial state to a given final state can be achieved with the help of quantum optimal control theory (QOCT). This tutorial provides an introduction to QOCT. It shows how the control equations defining such an optimal pulse follow from the variation of a properly defined functional. We explain the most successful schemes to solve these control equations and show how to incorporate additional constraints in the pulse design. The algorithms are then applied to simple quantum systems and the obtained pulses are analyzed. Besides the traditional final-time control methods, the tutorial also presents an algorithm and an example to handle time-dependent control targets.
\end{abstract}

\submitto{\JOB}
\section{Introduction}
\label{sec:ntro}
Since the first realization of the laser by T.H. Maiman in 1960,
physicists and chemists have had the vision to coherently control quantum systems using laser fields. For example, laser pulses may be applied to create and break a particular bond in a molecule, to control charge transfer within molecules, or to optimize high harmonic generation.
%
The first approach to break a certain bond in a molecule using a laser tuned on resonance and initiating a resonance catastrophe failed \cite{RVMK2000}. The molecule internally converted the energy too quickly, so that the specific bond did not break but instead the whole molecule was ``heated'' \cite{WRD93}. To overcome this so-called internal vibrational relaxation (IVR) a smarter excitation strategy and further technological improvements were necessary.

With the advent of femtosecond laser pulses in the 1980's and a sophisticated pulse-shaping technology \cite{WLPW92} the goal of controlling complex chemical reactions with coherent light was finally achieved: For example, in 1998 Assion {\it et al.}\ \cite{A98} showed that the product ratio \chem{CpFeCOCl^+}/\chem{FeCl^+} of the organo-metallic compound (\chem{CpFe(CO)_2Cl}) can be either maximized or minimized by a specially tailored light pulse; or in 2001, Levis {\it et al.}\ \cite{LMR2001} demonstrated a rearrangement of molecular fragments. 
In both of these experiments adaptive laser pulse shaping techniques \cite{WLPW92,JR92} were applied, i.e., a computer analyzes the outcome of the experiment and modifies the laser pulse shape to optimize the yield of a predefined reaction product. This process is repeated until the optimal laser pulse is found (see \aref{sec:cllexperiments}). The number of experiments based on this so-called closed-learning-loop (CLL) is growing constantly, see for example Refs.~\cite{B97,B2000,BDNG2001,HWCZM2002,BDKG2003,D2003,PWWSG2005}. Recently, the pulse shaping techniques have been extended to allow also for polarization shaping \cite{VKNNG2005, POS2006, PWWL2006a, PWWL2006b}, i.e., experimentalists can independently shape polarization, amplitude, and phase. 

In addition to further technological advances, it is of utmost importance to have powerful theoretical methods available. 
The questions that theory has to answer can be divided into two classes: 
The first class is that of {\it controllability} \cite{HTC83}, or in other words: Given a certain quantum system (e.g.\ a molecule) can the control target (a certain reaction product) be reached at all with the given controller (e.g.\ a laser)? 
The second class concerns the problem of finding the best way to achieve a given control objective, e.g., calculating the optimal laser pulse for breaking a particular bond in a molecule. 
%
%
Such theoretical predictions are very important to gain insight into the complexity of the control process, to determine the experimental parameters, to transfer laser pulse designs to the laboratory, or to compare the optimized pulse from an experiment with the calculated one \cite{B2004}. 

Several theoretical approaches have been developed to optimize laser pulses, ranging from brute-force optimization of a few pulse parameters \cite{JJMS90}, pulse-timing control \cite{TR86,TKR86,MHPB2003}, Brumer-Shapiro coherent control \cite{BS86}, stimulated-Raman-Adiabatic-Passage (STIRAP) \cite{GRSB90,CB92,BTS98}, to genetic algorithms \cite{CC2001}.
The most powerful approach, in our opinion, is optimal control theory (OCT) which is commonly applied in engineering, for example to design trajectories for satellites and space probes. 
%
The application of optimal control theory to quantum mechanics started in the late 1980's \cite{K89,PDR88} and shows continuous advances until today. Among the most important developments were the introduction of rapidly converging iteration schemes \cite{TKO92,ZBR98,ZR98,MT2003}, the generalization to include dissipation (Liouville space) \cite{AKT93,AKT97,OZR99}, and to account for multiple control objectives \cite{O2001}.

This tutorial will focus on the theoretical aspects of quantum optimal control theory (QOCT) and tries to explain the beginning Phd student how to calculate optimal laser pulses. The first step is to work through the basic theory which is presented in \sref{sec:theory}. 
In the following sections we will always refer to parts of the presented theory therein. The application of the derived method is then explored with the help of a two-level system in \sref{sec:tls_example}. The basics of the two-level system are reviewed in \aref{sec:TLStheory}. As a second example we consider the control of 
a 1D asymmetric double well model in \sref{sec:adw_example} which will motivate the need for further constraints on the optimal laser field. Theory and algorithms for optimizing laser pulses with additional constraints are explained in \sref{sec:ftheory}. Two examples for the asymmetric double well can be found in \sref{sec:direct} and \sref{sec:indirect031}.
The last part of this tutorial in \sref{sec:theoryextent} will focus on time-dependent control targets. A brief summary and outlook of this article can be found in \sref{sec:conclusion}.

We would like to conclude this introduction by emphasizing that our intention is to provide only a brief overview, but a detailed derivation of the algorithms, and simple examples that can be reworked by the reader (especially those of the two-level system).
The tutorial cannot cover all topics in quantum control and excludes the following imported topics: Closed-loop control (see Refs.~\cite{L2000}), time optimal control (see Refs.~\cite{W2000, KBS2001, WLW2002}), and dissipative systems (see Refs.~\cite{AKT93,AKT97,OZR99,PJ2005,BOS2006}).
For an excellent review on the experimental aspects of quantum control we like to refer the reader to Ref.~\cite{WB2001}.

\section{Theory and Algorithms}
\label{sec:theory}
%
In this section we sketch the basics of optimal control theory applied to quantum mechanics.

Let us consider an electron in an external potential $V({\bf r})$ under the influence of a laser field propagating in $z$-direction. Given an initial state $\Psi({ \bf r },0)=\phi({ \bf r })$, the time evolution of the electron is described by the  time-dependent Schr\"odinger equation with the laser field modelled in the dipole approximation (length gauge)
\begin{eqnarray}
\rmi \frac{\partial}{\partial t}\Psi({ \bf r },t)&=&\widehat{H}\Psi({ \bf r },t),\label{1SE}\\
\widehat{H}&=&\widehat{H}_0-\hat{{\bmu}}{\bepsilon}(t),\\
\widehat{H}_0&=&\widehat{T}+\widehat{V},
\end{eqnarray}
(atomic units are used throughout: $\hbar = m =e =1 $). Here, $\hat{{\bmu}}=(\hat{\mu}_x,\hat{\mu}_y)  $ is the dipole operator, and ${\bepsilon}(t)=(\epsilon_x(t),\epsilon_y(t))$ is the time-dependent electric field. The kinetic energy operator is $\widehat{T}=-\frac{\nabla^2}{2}$.\\
%
\subsection{Controllability}
\label{sec:completecontrol}
%
Before trying to find an optimal control for a given target and quantum system we raise the question if the given control target can be reached at all with the given controller, here: the controller is the laser field $\hat{H_1}=-\hat{\bmu} \bepsilon(t)$.  
In the following we want to summarize some of the rigorous results on controllability that exist in the literature. 

The most powerful and easy-to-use statements are available for $N$-level systems \cite{HTC83,RSDRP95,SFS2001}. We start with a definition of the term complete controllability and then discuss the results from Ref.\ \cite{RSDRP95}. 

{\it Definition (Schirmer et al.\ \cite{SFS2001}):} A quantum system 
\bea
\label{eq:Nlev_H}
\hat{H} &=& \hat{H}_0 + \hat{H}_I, \\
\label{eq:Nlev_H0I}
 \hat{H}_0 &=& \sum_{n=1}^{N} \varepsilon_n |n \ra \la n|, \qquad 
\hat{H}_I =  \sum_{m=1}^{M} f_m(t) \hat{H}_m,
\eea
is {\it completely controllable} if every unitary operator $\hat{U}$ is accessible from the identity operator $\hat{I}$ via a path $\gamma(t) = \hat{U}(t,t_0)$ that satisfies 
\bea
\rmi \partial_t \hat{U}(t,t_0) = \left(\hat{H}_0 + \hat{H}_I \right)\hat{U}(t,t_0) . 
\eea 

{\it Theorem (Ramakrishna et al.\ \cite{RSDRP95}):} A necessary and sufficient condition for complete controllability of a quantum system defined by equations \eref{eq:Nlev_H} and \eref{eq:Nlev_H0I} is that the Lie algebra $L_0$ has dimension $N^2$. 

We can therefore check the controllability of an $N$-level system by constructing its Lie algebra $L_0$ which is generated by $\hat{H}_0, \ldots,\hat{H}_M$ and then calculate the rank of the algebra. An algorithm for this task has been suggested in Ref.\ \cite{SFS2001}. This scheme is demonstrated for the two-level-system in \aref{sec:cc2level}. A further statement in Ref.\ \cite{RSDRP95} guarantees that complete controllability can also be achieved under external constraints on the strength of the controller. 

The extension of such controllability theorems to an infinite-dimensional Hilbert space and the inclusion of unbound operators like $\brr$ or $\nabla_{\brr}$ turns out to be non-trivial as shown in Ref. \ \cite{HTC83}. The conditions of controllability are only valid for quantum systems with a non-degenerate and discrete spectrum and do not include external constraints on the strength of the control functions.
%

\subsection{Derivation of the control equations}
\label{sec:stdoct}
%
Let us consider the following quantum mechanical control problem:

Our goal is to find a laser pulse $\bepsilon(t)$ which drives a quantum system from its initial state $\Psi(0)$ to a state $\Psi(T)$ in such a way that the expectation value of an operator $\widehat{O}$ is maximized at the end of the laser interaction:
\bea
\label{eq:j1}
\max_{\bepsilon(t)} {J_1} \qquad \mbox{with} \,\,\,\,  J_1[\Psi] = \left\langle  \Psi(T)| \hat{O} |\Psi(T) \right\rangle.
\eea

At this point we keep the operator $\hat{O}$ as general as possible. The only restriction on $\hat{O}$ is that it has to be a Hermitian operator. We will discuss examples for the operator $\hat{O}$ in \sref{sec:moperator}.
In addition to the maximization of $J_1[\Psi]$, we require that the fluence of the laser field is as small as possible which is cast in the following mathematical form:
\bea
\label{eq:penalty}
J_2[{ \bepsilon }] &=& - \sum_j  \int_0^T \!\! dt \,\, \alpha_j {  \epsilon_j }^2(t), \qquad j=x,y.  
\eea
where $\epsilon_x(t)$ and $\epsilon_y(t)$ are the components of the laser field perpendicular to the propagation direction. 
The positive constants $\alpha_j$ play the role of penalty factors: The higher the laser fluence the more negative the expression, the smaller the sum $J_1 + J_2$.
As pointed out in Ref.\ \cite{SV99}, the penalty factor can be extended to a \td function $\alpha_j(t)$ to enforce a given \td shape of the laser pulse, e.g. a Gaussian or sinusoidal envelope.
The constraint that the electronic wave function has to satisfy the time-dependent Schr\"odinger equation (TDSE) is expressed by
\begin{eqnarray}
\label{eq:j3}
  J_3[{ \bepsilon },\Psi,\chi]&=&
    - 2 \im \int_{0}^{T}\!\!  dt \,\, \left\langle\chi(t) \left| \left(\rmi \partial_t
    -\hat{H}(t)\right) \right| \Psi(t)\right\rangle,
\end{eqnarray} 
where we have introduced the Lagrange multiplier $\chi(t)$. Since we require the TDSE to be fullfilled by the complex conjugate of the wave-function as well, we obtain the imaginary part $\im$ of the functional. Note, that we choose the imaginary part to be consistent with the literature, for instance, Ref.\ \cite{PDR88, RSDRP95, ZBR98}.


The Lagrange functional has the form
\begin{eqnarray}
\label{eq:stdfunctional}
J[\chi,\Psi,{ \bepsilon }] = J_1[\Psi] + J_2[{ \bepsilon }] + J_3[\chi,\Psi,{ \bepsilon }].
\end{eqnarray} 
We refer to this functional as the standard optimal control problem and start the discussion of all extensions considered in this work from this standard form.
%
\subsection{Variation of $J$}
%
To find the optimal laser field from the functional in equation \eref{eq:stdfunctional} we perform a total variation. Since the variables $\Psi$, $\chi$ and $\epsilon$ are linearly independent we can write:
\bea 
\nonumber
\delta J &=&  \int_0^T \!\!  d\tau \!\! \int \!\! d \brr \,\left\{\frac{\delta J}{\delta \Psi(\brr,\tau)}\delta \Psi(\brr,\tau) + \frac{\delta J}{\delta \chi(\brr,\tau)}\delta \chi(\brr,\tau) \right\} + \sum_{k=x,y} \int_0^T \!\!  d\tau \, \frac{\delta J}{\delta \epsilon_k(\tau)}\delta \epsilon_k(\tau) \\
\label{eq:totalvarJ}
 &=& \delta_{\Psi} J +  \delta_{\chi}J +  \sum_{k=x,y} \delta_{\epsilon_k} J, 
\eea
where we have omitted the derivations with respect to the complex conjugate of the wave function $\Psi^*(t)$ and the Lagrange multiplier $\chi^*(t)$, since the functional derivative will result in the complex conjugate equations for these variations.

Since we are looking for a maximum of $J$, the necessary condition is 
\bea
\nonumber
&&\delta J = 0\\
\label{eq:nc1}
&&\Rightarrow \delta_{\Psi} J = 0, \qquad \delta_{\chi} J = 0, \qquad \delta_{\epsilon_k} J=0. 
\eea
\subsubsection{Variation with respect to the wave function $\Psi$}
First, let us consider the functional derivative of $J$ with respect to $\Psi$:
\bea
\nonumber
\frac{\delta J_1}{\delta \Psi(\brr',\tau)} &=& \, \hat{O}\Psi^*(\brr',\tau) \,\delta(T-\tau),\\
\nonumber
\frac{\delta J_2}{\delta \Psi(\brr',\tau)} &=& 0,\\
\label{eq:fderivJ}
\frac{\delta J_3}{\delta \Psi(\brr',\tau)} &=& -\rmi \left(\rmi \partial_\tau + \hat{H}(\tau) \right) \chi^*(\brr',\tau) - \left[\chi^*(\brr',t) \delta(t-\tau)\right]\Big|_0^T.
\eea
For the last functional derivative we have used the following partial integration:
\bea
\nonumber
& &\int_0^T \!\! dt \,\, \left\langle  \chi(t) | \rmi \partial_t - \hat{H}(t) | \Psi(t) \right\rangle  \\\nonumber
&& = \rmi  \left\langle  \chi(t)|\Psi(t) \right\rangle \Big|_0^T - \rmi \int_0^T \!\! dt \,\, \left\langle  \partial_t  \chi(t)| \Psi(t) \right\rangle  
- \int_0^T \!\! dt \,\, \left\langle \hat{H}(t)  \chi(t)| \Psi(t) \right\rangle \\\nonumber
&& = \rmi  \left\langle \chi(t)| \Psi(t) \right\rangle \Big|_0^T  + \int_0^T \!\! dt \,\,  \left\langle \left( \rmi \partial_t -  \hat{H}(t) \right) \chi(t) |\Psi(t) \right\rangle.  
\eea
%
We find for the variation with respect to $\Psi$:
\bea
\nonumber
\delta_{\Psi} J &=& \left\langle \Psi(T)|\hat{O}| \delta \Psi(T) \right\rangle + \rmi \int_0^T \!\! d\tau \left\langle \left( \rmi \partial_\tau - \hat{H}(\tau) \right) \chi(\tau) |\delta \Psi(\tau) \right\rangle \\
\label{eq:varpsi}
&& \qquad - \left\langle \chi(T) | \delta \Psi(T) \right\rangle + \underbrace{\left\langle \chi(0) | \delta \Psi(0) \right\rangle}_{=0} .
\eea
The variation of $\delta \Psi(0)$ vanishes because we have a fixed initial condition, $\Psi(0) = \phi_i$.
\subsubsection{Variation with respect to the Lagrange multiplier $\chi$} 
Now we do the same steps for $\chi(t)$:
\bea \nonumber
\frac{\delta J_1}{\delta \chi(\brr',\tau)} &=& \frac{\delta J_2}{\delta \chi(\brr',\tau)} = 0, \\\nonumber
\frac{\delta J_3}{\delta \chi(\brr',\tau)} &=&  \rmi \left(\rmi \partial_{\tau} + \hat{H}(\tau) \right) \Psi^*(\brr',\tau).
\eea
In contrast to the variation with respect to $\Psi$ we do not have boundary terms here. The variation of $J$ with respect to $\chi$ yields
\bea
\label{eq:varchi}
\delta_{\chi} J &=& - \rmi \int_0^T \!\! d\tau \left\langle \left(\rmi \partial\tau - \hat{H}(\tau) \right) \Psi(\tau) |\delta \chi(\tau) \right\rangle,  \qquad k=x,y.
\eea
\subsubsection{Variation with respect to the field}
The functional derivative with respect to $\epsilon_k(t)$ is
\bea\nonumber
\frac{\delta J_1}{\delta \epsilon_k(\tau)} &=& 0, \\\label{eq:dereps1}
\frac{\delta J_2}{\delta \epsilon_k(\tau)} &=& -2 \alpha_k \epsilon_k(\tau),\\\label{eq:dereps2}
\frac{\delta J_3}{\delta \epsilon_k(\tau)} &=& -2  \, \im  \left\langle \chi(\tau) | \hat{\mu}_k| \Psi(\tau) \right\rangle, \qquad k=x,y.
\eea
Hence, the variation with respect to $\epsilon_k(t)$ yields
\bea
\label{eq:vareps}
 \delta_{ \epsilon_k} J = \int_0^T \!\! d\tau \,\,\left\{ - 2 \im \left\langle \chi(\tau) | \hat{\mu}_k |\Psi(\tau) \right\rangle - 2 \alpha_k \epsilon_k(\tau) \right\} \delta \epsilon_k(\tau).
\eea

%
\subsection{Control equations}
%
Setting each of the variations independently to zero results in the desired
 control equations.
From $\delta_{\epsilon_k} J=0$ [equation \eref{eq:vareps}] we find
\bea
\label{eq:ctrleps}
 \alpha_k \epsilon_k(t) = -\im\,\langle\chi(t)|\hat{\mu}_k|\Psi(t)\rangle,  \qquad k=x,y. 
\eea
The laser field $\bepsilon(t)$ is calculated from the wave function $\Psi(t)$ and the Lagrange multiplier $\chi(t)$ at the same point in time.
The variation  $\delta_{\chi} J $ in equation \eref{eq:varchi} yields a \TDSE~for $\Psi(t)$ with a fixed initial state $\phi_i$,
\bea
\label{eq:SE}
0 &=& \left( \rmi \partial_t - \hat{H}(t) \right) \Psi({ \bf r },t), \qquad 
\Psi({ \bf r },0) = \phi_i({ \bf r }).
\eea
Note that this equation also depends on the laser field $\bepsilon(t)$ via the Hamiltonian.

The variation with respect to the wave function $\delta_{\Psi} J$ in equation \eref{eq:varpsi} results in
\bea
\label{eq:SE1}
\left(\rmi \partial_t - \hat{H}(t)\right)\chi({ \bf r },t) =
 \rmi \left(\chi({ \bf r },t)-\hat{O} \Psi({ \bf r },t)\right)\delta(t-T).
\eea
If we require the Lagrange multiplier $\chi(t)$ to be continuous at $t=T$, we can solve the following two equations instead of equation \eref{eq:SE1}:
\begin{eqnarray}
\label{eq:SE2}
 \left(\rmi \partial_t - \hat{H}(t)\right)\chi({ \bf r },t)&=&0,\\
\label{eq:SE3}
\chi({ \bf r },T) &=&\hat{O} \Psi({ \bf r },T).
\end{eqnarray}
To show this we integrate over equation \eref{eq:SE1}:
\begin{eqnarray}
\nonumber
&& \lim_{\kappa\to 0}\int_{T-\kappa}^{T+\kappa}\!\!\!\!\!\!dt\left[\left(\rmi \partial_t - \hat{H}(t)\right)\chi({ \bf r },t) \right]\\
\label{eq:proof1}
&=&\lim_{\kappa\to 0}\int_{T-\kappa}^{T+\kappa}\!\!\!\!\!\!dt\; \rmi \left(\chi({ \bf r },t)-\hat{O} \Psi({ \bf r },t)\right)\delta(t-T).
\end{eqnarray}
The left-hand side of equation \eref{eq:proof1} vanishes because the integrand is a continuous function. It follows that also the right-hand side must vanish, which implies equation \eref{eq:SE3}. From equations \eref{eq:SE3} and \eref{eq:SE1} then follows equation \eref{eq:SE2}.
Hence, the Lagrange multiplier $\chi(t)$ satisfies a time-dependent Schr\"odinger equation with an initial condition at $t=T$. 
The set of equations that we need to solve is now complete:  \eref{eq:ctrleps},  \eref{eq:SE}, \eref{eq:SE2}, and \eref{eq:SE3}. 

%
%
To find an optimal field $\bepsilon(t)$ from these equations we use an iterative algorithm which will be discussed in the \sref{sec:malgorithm}. 
\subsection{Target operators}
\label{sec:moperator}
%
So far we have shown how we can optimize a laser field and which equations have to be solved to achieve this goal. Before we discuss details on how the equations are solved in practice, we want to discuss different examples for the target operator $\hat{O}$. 

\subsubsection{Projection operator}
%
Choosing a projection operator $\hat{O}=| \phi_f \rangle \langle \phi_f |$, the maximization of $J_1$ corresponds to maximizing the overlap of the propagated wave function $\Psi(T)$ with $\phi_f$, i.e.,
\bea
\label{eq:J1proj}
J_1 = \left\langle \Psi(T) | \phi_f \right\rangle \left\langle \phi_f | \Psi(T)\right\rangle = 
\left|\left\langle \Psi(T) | \phi_f \right\rangle \right|^2.
\eea 
Using a projection operator as target in the optimal control algorithm therefore allows one to find an optimized pulse which drives the system from the initial state $\Psi(0)$ to the desired target state $\phi_f$ up to a global phase factor $\rme^{\rmi \gamma}$. In other words, the target functional $J_1$ is invariant under the transformation
\bea
\phi_f \to  \rme^{\rmi \gamma} \, \phi_f .
\eea
It is possible to fix this phase if we replace equation \eref{eq:J1proj} by the following functional \cite{TR2001}:
\bea
\label{eq:j1re}
\re \left\langle \Psi(T) | \phi_f \right\rangle ,
\eea
which can be derived in the following way:
\bea
\label{eq:mintarget}
\mathrm{min} \, \| \Psi(T) - \phi_f  \|^2 = \mathrm{min} \, \left\{ \left\langle \Psi(T) | \Psi(T) \right\rangle + \left\langle \phi_f | \phi_f \right\rangle - 2 \re \left\langle \Psi(T) | \phi_f \right\rangle \right\}.
\eea
Assuming normalized states the minimization corresponds to the maximization of equation \eref{eq:j1re}. 

In all cases considered in this work, $\phi_f$ is chosen to be an excited state of the quantum system. But it is also possible to choose any normalized superposition of bound and continuum states as a target state. For example, choosing
\bea
\phi_f = N_\gamma \exp[-\gamma (\brr-\brr_0)^2] \rme^{\rmi {\bf k}_0 \brr}
\eea
as target state allows us to find a laser field that drives the particle to a predefined expectation value for position and momentum: $\langle \phi_f |\hat{\brr}| \phi_f \rangle=\brr_0$ and $\langle \phi_f| \hat{\bf p}| \phi_f \rangle={\bf k}_0$. This choice is of course not unique since, for example, any choice for the width $\gamma$ results in the same expectation values. 

\subsubsection{Local operator}
\label{sec:locop}
%
Instead of using a non-local operator like the projection operator we can also employ a local operator for $\hat{O}$, e.g., $\hat{O}=f(\brr)$.
The most popular example for the local operator is the $\delta$~function,
\bea
J_1 = \left\langle \Psi(T) |\delta(\brr-\brr_0)| \Psi(T) \right\rangle =  |\Psi(T,\brr_0)|^2,
\eea
which maximizes (in the single-particle case) the density at point $\brr_0$. The more density is squeezed into the point the higher the yield $J_1$. The target by itself does not sound very physical but in practice it allows us to maximize the density distribution in a given region in space. In the next section we will show that a function different from the $\delta$~function corresponds to a multi-objective optimization.
In the practical implementation the $\delta$~function will be approximated by a narrow Gaussian.

A different choice for driving the density towards a target density $n_f$ [similar to equation \eref{eq:mintarget}] is
\bea
\label{eq:mintarget2}
\mathrm{min} \, \| \sqrt{n(T)} - \sqrt{n_f}  \|^2 =\mathrm{min} \, \left\{ 2 - 2\int \!\! d\brr \, \sqrt{n(\brr, T) \, n_f(\brr)} \right\},
\eea
where we assume normalization for $n(T)$ and $n_f$.
The minimization corresponds to a maximization of
\bea
\label{eq:denssqr}
\tilde{J}_1 = \int \!\! d\brr \, \sqrt{n(\brr,T) \,n_f(\brr)}.
\eea
\subsubsection{Multi-objective target operators}
%
Within the formalism established, we are not restricted to a single objective. We can also employ a multi-objective target operator like
\bea
\hat{O} = \sum_j \beta_j \hat{O}_j.
\eea
 For example, the operators $\hat{O}_j$ can be projection operators for different excited states. The weights $\beta_j$ are chosen to balance the different objectives. If $\beta_j$ is chosen negative, the optimization will try to minimize the expectation value of $\hat{O}_j$, e.g., the occupation of a specific excited state. Combinations of projection and local operators are also possible.
Note that the choice of a sum of projection operators has to be distinguished from the projection on the coherent superposition of these states. 

A local operator which is not a $\delta$~function corresponds to a multi-objective optimization. This can be shown by considering the limit of infinitely many $\delta$~target operators
\bea
\hat{O}(\brr) &=& \int \!\! d\brr' \,\, \beta(\brr) \delta(\brr'-\brr), \\
J_1 &=& \int \!\! d\brr \, \beta(\brr) |\Psi(\brr,T)|^2 ,
\eea
where the weight function $\beta(\brr)$ can be identified with an arbitrary local operator. 
\subsubsection{Finite penalty versus complete controllability}
%
Introducing a (positive) penalty factor $\balpha$ has the immediate consequence that a final target state occupation of $100\%$  cannot be achieved. This can be proven within a few steps by achieving a contradiction:

 Assume that we have found the optimal field $\bepsilon_{\mathrm{opt}}(t)$ which drives the system from the initial state $\phi_i = \Psi(0)$ to the target state $\Psi(T) = \phi_f$. According to equation \eref{eq:SE3} the initial state for the Lagrange multiplier is then $\chi(T)=\phi_f$. Since the two Hamiltonians of equations \eref{eq:SE} and \eref{eq:SE2} are the same, the time-evolution operators for $\Psi(t)$ and $\chi(t)$ are identical:
\bea
\nonumber
&&\Psi(T) = \hat{U}(T,t) \hat{U}(t,0) \Psi(0) = \chi(T) \\
\nonumber
&& \Rightarrow \hat{U}(t,0) \Psi(0) = \hat{U}(t,T)\chi(T) \\
\nonumber
&& \Rightarrow  \Psi(t) = \chi(t).
\eea
Inserting this finding in equation \eref{eq:ctrleps} gives
\bea
\nonumber
\alpha_k \epsilon_k(t) = -\im\langle\Psi(t)|\hat{\mu}_k|\Psi(t)\rangle =0 ,  \qquad k=x,y,
\eea
resulting in the statement that $\bepsilon_{\mathrm{opt}}(t)=0$ or that $100 \%$ overlap cannot be achieved. 
Except for the trivial case $\bepsilon(t) =0 $ which presents a minimum for the functional if the initial and target state are orthogonal, $\langle \phi_i | \phi_f \rangle = 0$, we have to deal with the fact that $100\%$ occupation of the final state cannot be obtained with this kind of algorithm even if the system is {\it completely controllable} in principle. 

\subsection{Algorithm to solve the control equations}
\label{sec:malgorithm}
In the following we present two approaches to solve the standard optimal control problem [see equations \eref{eq:j1}, \eref{eq:penalty}, and \eref{eq:j3}]. The first approach is an iterative solution \cite{ZR98} of the equations \eref{eq:ctrleps}, \eref{eq:SE}, \eref{eq:SE2}, and \eref{eq:SE3}. It provides the starting point for all extensions developed in this work. The second scheme applies to the case where the target operator is a projection operator. In this case a slightly faster algorithm can be deduced.

As the word {\it iterative} already indicates, it will be necessary to solve the \TDSE~more than once. Even with the present computational resources this limits the application of the algorithms to relatively low-dimensional systems.


\subsubsection{Standard iterative scheme}
\label{sec:malgorithm_iter}
%
The control equations (\ref{eq:ctrleps}), (\ref{eq:SE}), (\ref{eq:SE2}), and (\ref{eq:SE3}) can be solved as follows \cite{ZR98}.
The scheme starts with propagating $\phi_i=\Psi^{(0)}(0)$ forward in time. For the initial propagation we have to guess the laser field $\bepsilon^{(0)}(t)$. For most of the cases we consider, the trivial initial guess $\bepsilon^{(0)}(t)=0$ is sufficient. However, sometimes the algorithm gets stuck in this solution. In this case, a rule of thumb for the initial choice is that the forward propagation has to result in a large enough value for $\| \hat{O} \Psi(T)\|^2$. A strong dc-field $\bepsilon^{(0)}(t)=const$ often proves to be helpful. 
This part of the iteration can be expressed symbolically by
\begin{equation}
\nonumber
\begin{array}{l c c c c c c l }
  {\mbox{step 0:}} \,\, & \Psi^{(0)}(0) & \overset{{\bepsilon}^{(0)}(t)}{\longrightarrow} &
 \Psi^{(0)}(T). &  &  &  &
\end{array}
%
%
%
%
\end{equation}

After the initial propagation we determine the final state for the Lagrange multiplier wave function $\chi^{(0)}(T)$ by applying the target operator to the final state of the wave function, $\hat{O} \Psi^{(0)}(T)$. The laser field for the backward propagation for $\chi^{(0)}(t)$, $\widetilde{\bepsilon}^{(0)}(t)$ is determined by 
\bea
\label{eq:fieldtilde}
\widetilde{\epsilon}_j^{(k)}(t) &=& - \frac{1}{\alpha_j}\im \left\langle \chi^{(k)}(t) |\hat{\mu}_j | \Psi^{(k)}(t) \right\rangle, \qquad j=x,y.
\eea
The propagation from  $\chi^{(0)}(T)$ to $\chi^{(0)}(T-dt)$ is done with the field $\widetilde{\bepsilon}^{(0)}(T)$, where we use $\chi^{(0)}(T)$ and $\Psi^{(0)}(T)$ in equation \eref{eq:fieldtilde}. The small error introduced here is compensated by choosing a sufficiently small time step. 
In parallel, we propagate $\Psi^{(0)}(T)$ backward with the previous field $\bepsilon^{(0)}(t)$. This additional parallel propagation is only necessary if the storage of $\Psi^{(0)}(t)$ in the memory is not possible. For the next propagation step from $\chi^{(0)}(T-dt)$ to $\chi^{(0)}(T-2\,dt)$ we use $\Psi^{(0)}(T-dt)$ and $\chi^{(0)}(T-dt)$ in equation \eref{eq:fieldtilde}. We repeat these steps until $\chi^{(0)}(0)$ is reached. To check the reliability of the parallel propagation we project $\Psi^{(0)}(0)$ onto $\phi_i$ and compare with $1$. We summarize the whole iteration step by
\begin{equation}
\label{eq:stepk1}
\begin{array}{l c c c c c c l }
   {\mbox{step k:}} \,\,   & & & \left[ \Psi^{(k)}(T) \right. & \overset{{\bepsilon}^{(k)}(t)}{\longrightarrow} &
 \left. \Psi^{(k)}(0) \right] & & \\
                   & & & \hat{O} \Psi^{(k)}(T) = \chi^{(k)}(T) & \overset{\widetilde{{\bepsilon}}^{(k)}(t) 
}{\longrightarrow} & \chi^{(k)}(0). & &
\normalsize
\end{array}
\end{equation}
%
The last part of the zeroth iteration step consists in setting $\Psi^{(1)}(0)=\phi_i$ and propagating $\Psi^{(1)}(0)$ forward with the field $\bepsilon^{(1)}(t)$ determined by 
\bea
\label{eq:fieldnext}
\epsilon^{(k+1)}_j(t) &=& - \frac{1}{\alpha_j} \im \left\langle \chi^{(k)}(t) |\hat{\mu}_j | \Psi^{(k+1)}(t) \right\rangle, \qquad j=x,y,
\eea
which requires the input of $\chi^{(0)}(t)$. Again, we have to use the saved values from the backward propagation or propagate from $\chi^{(0)}(0)$ to  $\chi(T)$ in parallel using the previously calculated field $\widetilde{\bepsilon}^{(0)}(t)$. We end up having calculated $\bepsilon^{(1)}(t)$ and $\Psi^{(1)}(T)$ which can be expressed by
\begin{equation}
\label{eq:stepk2}
\begin{array}{l c c c c c c l }

& & & &  & \left[ \chi^{(k)}(0) \right. & \overset{{\widetilde{\bepsilon}}^{(k)}(t)}{\longrightarrow} &
 \left. \chi^{(k)}(T) \right] \\
& & &  & & \phi_i = \Psi^{(k+1)}(0) & \overset{\bepsilon^{(k+1)}(t) 
}{\longrightarrow} & \Psi^{(k+1)}(T).
\normalsize
\end{array}
\end{equation}
This completes the zeroth iteration step. The loop is closed by continuing with equation \eref{eq:stepk1}, i.e., propagating $\hat{O} \Psi^{(1)}(T) = \chi^{(1)}(T)$ with $\widetilde{\bepsilon}^{(1)}(t)$ [see equation \eref{eq:fieldtilde}] backwards to $\chi^{(1)}(0)$.

If the initial guess for the laser field is appropriate the algorithm starts converging very rapidly and in a monotonic way, meaning that the value for the functional $J$ in equation \eref{eq:stdfunctional} is increasing at each iteration step. The monotonic convergence can be proven analytically \cite{ZBR98,ZR98}. In the proof an infinitely accurate solution of the \TDSE~is assumed. Since this is not possible in practice, it may happen that the functional decreases in the numerical scheme, e.g., when absorbing boundaries are employed. This sensitivity provides an additional check on the accuracy of the propagation. We can summarize the complete scheme by
%
%
%
\begin{equation}
\label{eq:scheme}
\scriptsize
\begin{array}{l c c c c c c l }
  {\mbox{step 0:}} \,\, & \Psi^{(0)}(0) & \overset{{\bepsilon}^{(0)}(t)}{\longrightarrow} &
 \Psi^{(0)}(T) &  &  &  &\\
   {\mbox{step k:}} \,\,   & & & \left[ \Psi^{(k)}(T) \right. & \overset{{\bepsilon}^{(k)}(t)}{\longrightarrow} &
 \left. \Psi^{(k)}(0) \right] & & \\
                   & & & \hat{O} \Psi^{(k)}(T) = \chi^{(k)}(T) & \overset{\widetilde{{\bepsilon}}^{(k)}(t) 
}{\longrightarrow} & \chi^{(k)}(0) & &\\
& & & &  & \left[ \chi^{(k)}(0) \right. & \overset{{\widetilde{\bepsilon}}^{(k)}(t)}{\longrightarrow} &
 \left. \chi^{(k)}(T) \right] \\
& & &  & & \phi_i = \Psi^{(k+1)}(0) & \overset{\bepsilon^{(k+1)}(t) 
}{\longrightarrow} & \Psi^{(k+1)}(T).
\end{array}
\normalsize
\end{equation}

\subsubsection{Projection operator - rapidly convergent scheme}
%
If the target operator is a projection operator $\hat{O}=|\phi_f \rangle  \langle \phi_f |$, a scheme with even faster convergence can be derived from the modified functional
\bea
\tilde{J}_3[{ \bepsilon },\Psi,\chi] = - 2 \im \left\{ \left\langle \Psi(T)| \phi_f \right\rangle \int_0^T\!\! dt \,\, \left\langle \chi(t) \left| \left(\rmi \partial_t
    -\hat{H}(t)\right) \right| \Psi(t) \right\rangle \right\}.
\eea 
The scheme differs only in two points. First, the iteration is started with propagating $\chi(t)$ backwards in time which is possible if we set $|\chi(T) \rangle=|\phi_f \rangle$. Second, the equations which determine the laser field [equations \eref{eq:fieldtilde} and \eref{eq:fieldnext}] are replaced by 
\bea
\label{eq:fieldtildeZBR98}
\widetilde{\epsilon}_j^{(k)}(t) &=& - \frac{1}{\alpha_j} \im \left\{\, \left\langle \Psi^{(k)}(t)| \chi^{(k)}(t)\right\rangle \, \left\langle \chi^{(k)}(t) |\hat{\mu}_j | \Psi^{(k)}(t) \right\rangle \, \right\}, \\
\label{eq:fieldnextZBR98}
\epsilon_j^{(k+1)}(t) &=& -\frac{1}{\alpha_j} \im \left\{ \, \left\langle \Psi^{(k+1)}(t)| \chi^{(k)}(t) \right\rangle\, \left\langle \chi^{(k)}(t) |\hat{\mu}_j | \Psi^{(k+1)}(t) \right\rangle\, \right\}, \,\,\,\, j=x,y.\qquad 
\eea
We can summarize this scheme by
\begin{equation}
\label{eq:schemeZBR98}
\scriptsize
\begin{array}{l c c c c c c l }
  {\mbox{step 0:}} \,\, & \phi_f = \chi^{(0)}(T) & \overset{{\bepsilon}^{(0)}(t)}{\longrightarrow} &
 \chi^{(0)}(0) &  &  &  &\\
   {\mbox{step k:}} \,\,   & & & \left[ \chi^{(k)}(0) \right. & \overset{{\bepsilon}^{(k)}(t)}{\longrightarrow} &
 \left. \chi^{(k)}(T) \right] & & \\
                   & & & \phi_i=\Psi^{(k)}(0) & \overset{\widetilde{{\bepsilon}}^{(k)}(t) 
}{\longrightarrow} & \Psi^{(k)}(T) & &\\
& & & &  & \left[ \Psi^{(k)}(T) \right. & \overset{{\widetilde{\bepsilon}}^{(k)}(t)}{\longrightarrow} &
 \left. \Psi^{(k)}(0) \right] \\
& & &  & & \phi_f = \chi^{(k+1)}(T) & \overset{\bepsilon^{(k+1)}(t) 
}{\longrightarrow} & \chi^{(k+1)}(0).
\end{array}
\normalsize
\end{equation}
Again, monotonic convergence can be proven. The scheme contains some freedom in the choice of the first overlap $\langle \Psi^{(k')}(t)| \chi^{(k)}(t)\rangle$ (with $k'=k$ or $k'=k+1$) in equations \eref{eq:fieldtildeZBR98} and \eref{eq:fieldnextZBR98}, since 
\bea
\nonumber
\left\langle \Psi(t)| \chi(t) \right\rangle = \left\langle \Psi(t')| \chi(t') \right\rangle.
\eea
The authors of Ref.\ \cite{ZBR98} report that the convergence changes for different choices of $\langle \Psi(t)| \chi(t)\rangle$. In our implementation we update this overlap at every point in time.

\subsection{Example: Two-level system}
\label{sec:tls_example}
%
Let us apply the previously developed theory to 
a two-level system. A brief review about the theory of two-level systems can be found in \aref{sec:TLStheory}. 
For the two-level system the integrals which determine the field, i.e., equations \eref{eq:fieldtildeZBR98} and \eref{eq:fieldnextZBR98}, reduce to
\bea
\nonumber
\la \Psi(t) | \chi(t) \ra &=& g_a^*(t) h_a(t) + g_b^*(t) h_b(t) , \\
\nonumber
\la \chi(t) |\hat{\mu}| \Psi(t) \ra &=& \rho_{ab} h_a^*(t) g_b(t) \rme^{\rmi (\omega_a - \omega_b) t} + \rho_{ba} h_b^*(t) g_a(t) \rme^{\rmi (\omega_b - \omega_a) t}\\ 
 &=& \mu \left( h_a^*(t) g_b(t) \rme^{-\rmi \omega_{ba} t} + h_b^*(t) g_a(t) \rme^{\rmi \omega_{ba} t} \right),
\eea
with $c_k(t)=\la k | \Psi(t)\ra $ and where $g_k(t) = c_k(t) \rme^{\rmi \omega_k t}$ was defined in equations \eref{eq:ode_ga_general} and \eref{eq:ode_gb_general}.  The coefficients of the Lagrange multiplier wave function $\chi(t)$ in the basis $|a \ra$ and $|b \ra$ are $\la k | \chi(t)\ra=l_k(t)= h_k(t) \rme^{-\rmi \omega_k t} $.

\begin{figure}[!h]
\subfigure[]{
  \label{fig:2levopt1}
  \resizebox{0.46\textwidth}{.26\textheight}{\includegraphics*{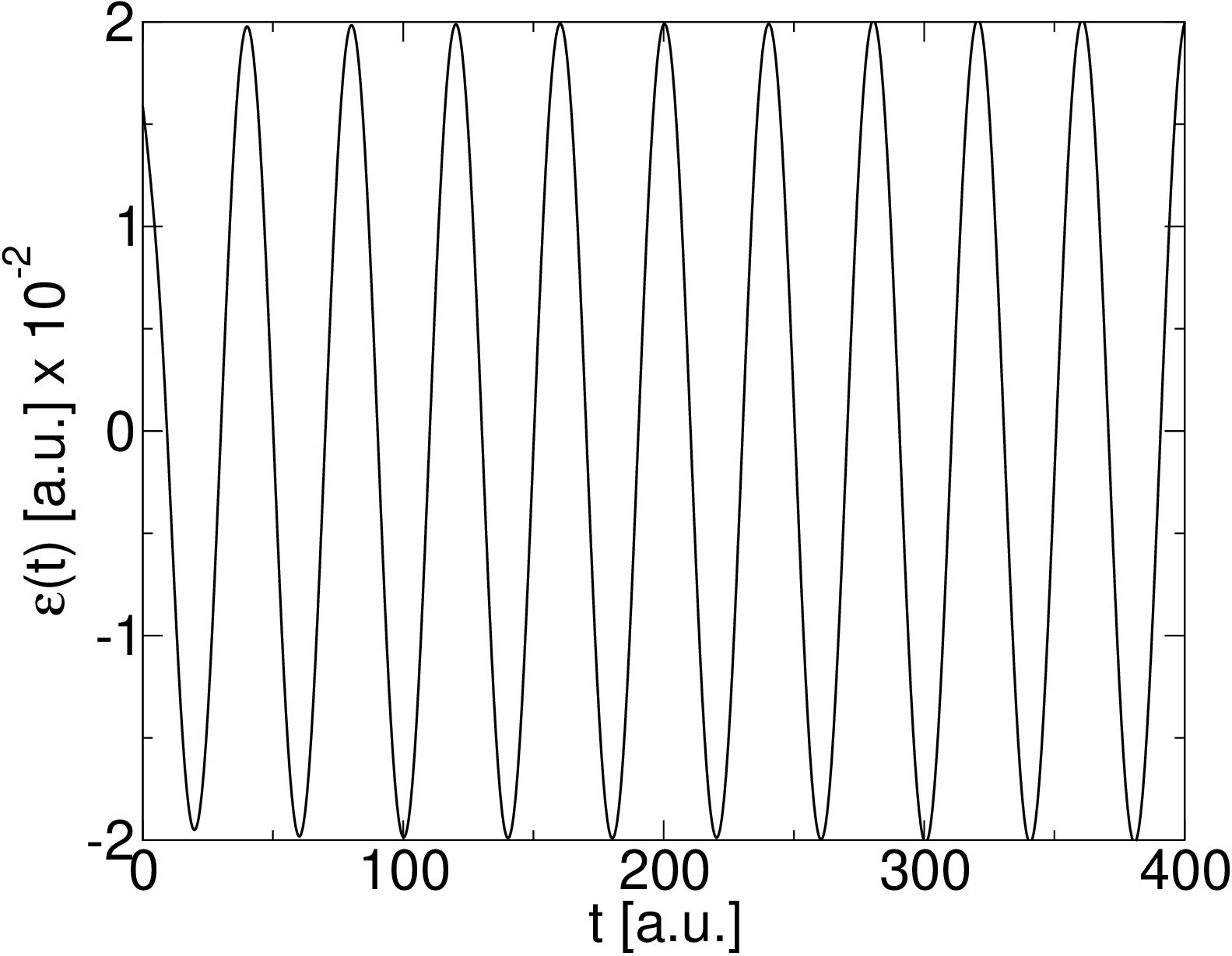}}
    }
\subfigure[]{
  \label{fig:2levopt2}
  \resizebox{0.46\textwidth}{.26\textheight}{\includegraphics*{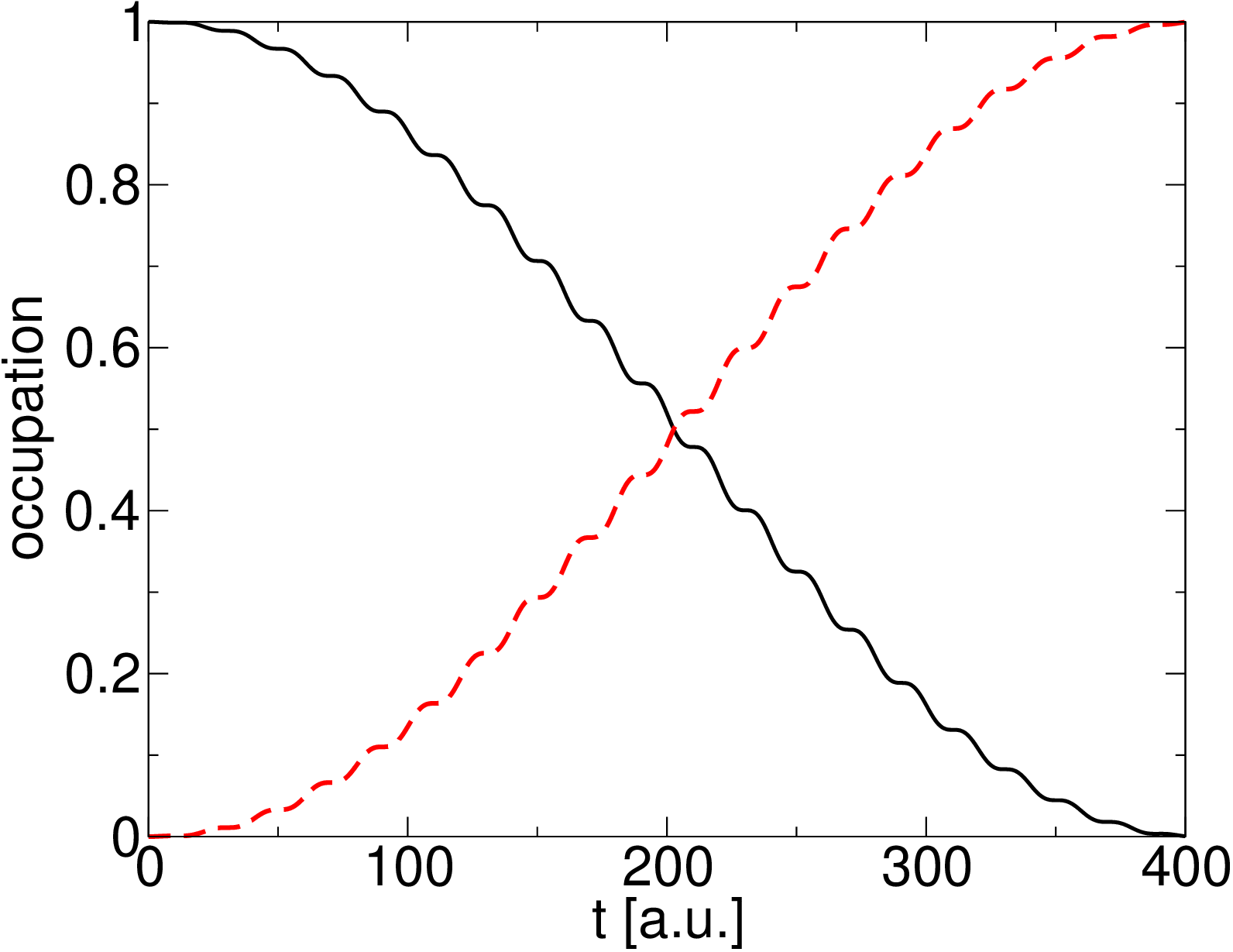}}
}\\
\subfigure[]{
  \label{fig:2levopt3}
  \resizebox{0.46\textwidth}{.26\textheight}{\includegraphics*{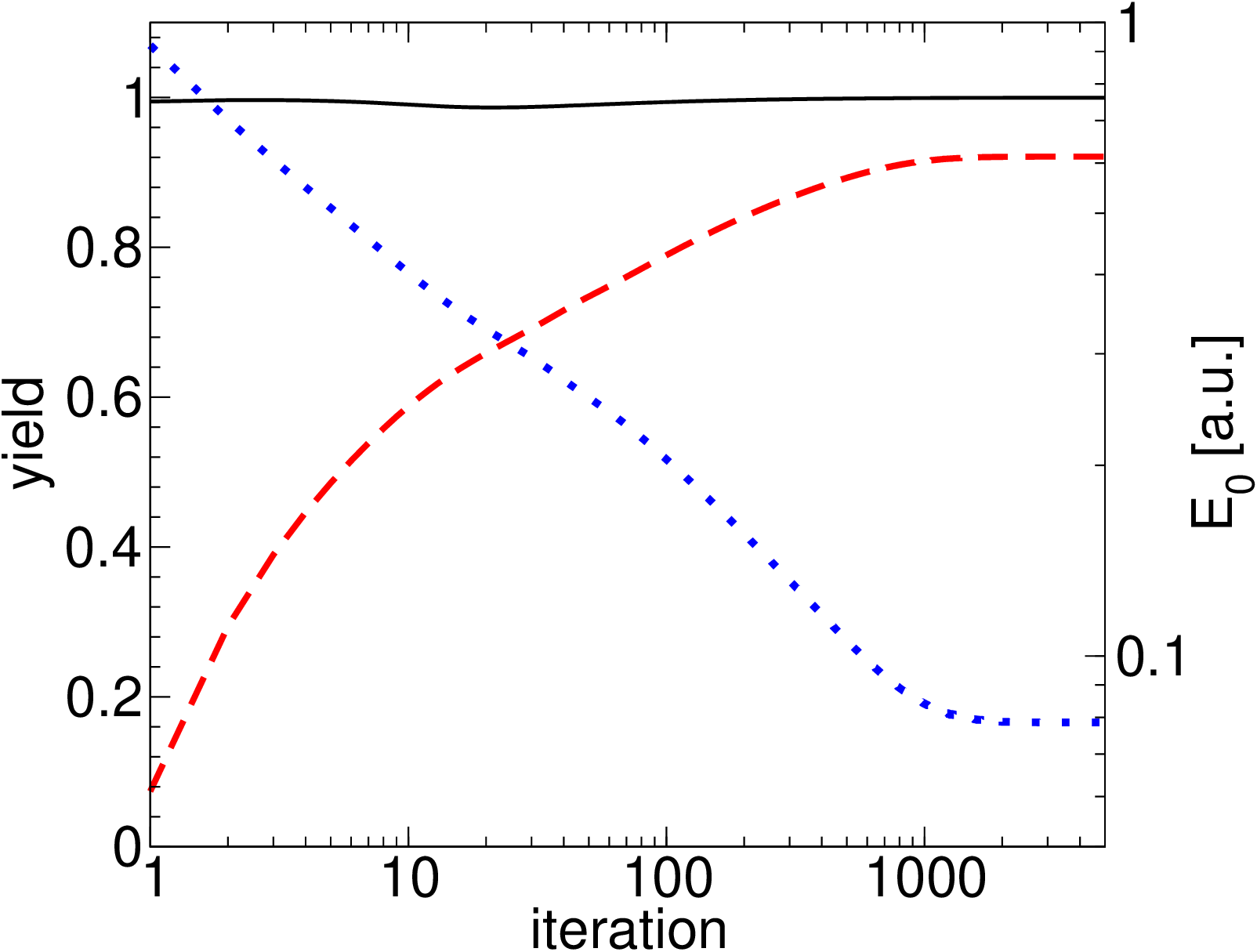}}
}
\subfigure[]{
  \label{fig:2levoptscan}
  \resizebox{0.46\textwidth}{.26\textheight}{\includegraphics*{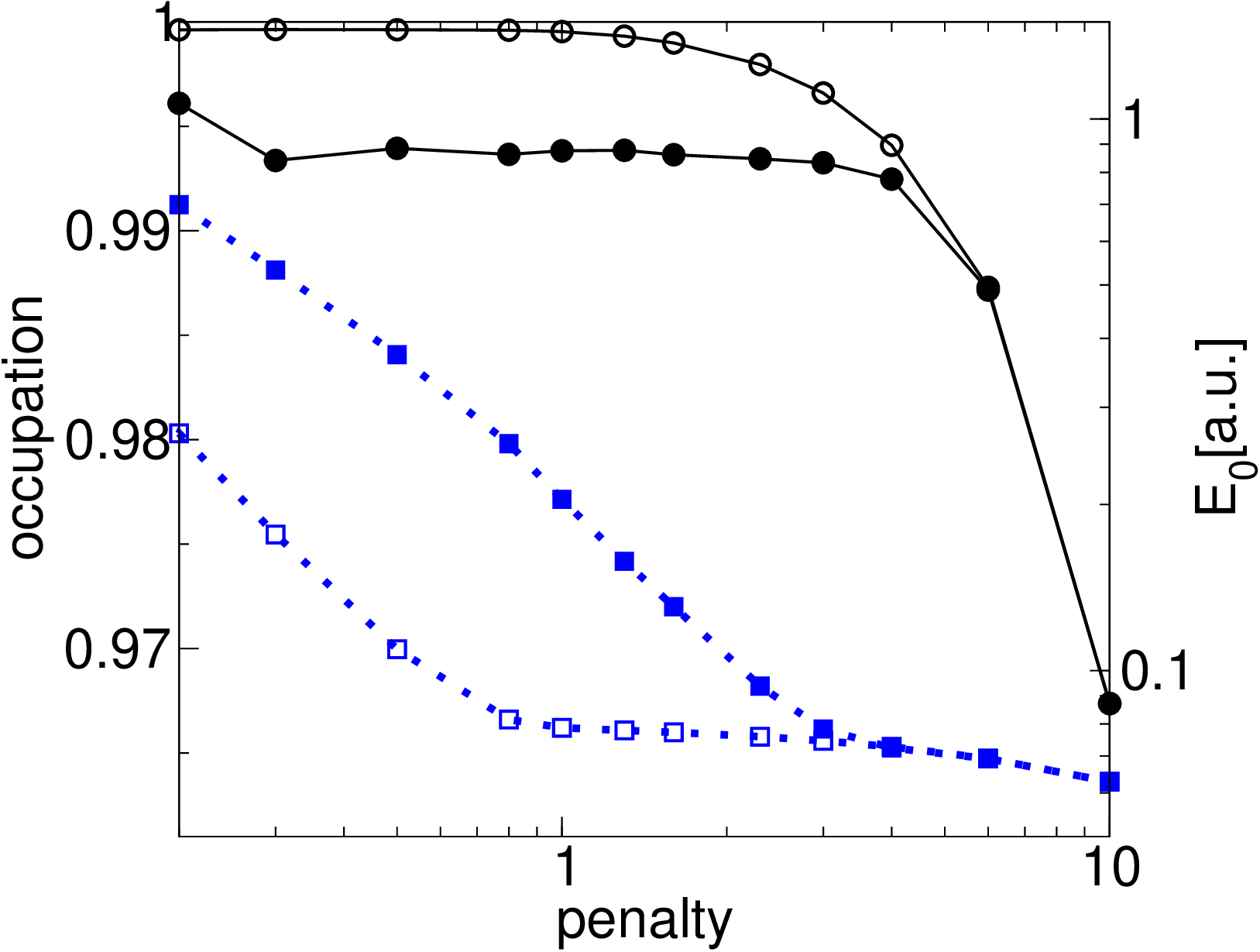}}
}
\caption[Optimal control for the two-level system.]{(Color online). We apply optimal control theory to invert the population of a two-level system. (a) Optimized laser field of the $5000$th iteration with $\alpha=1.0$. (b) Time evolution of the occupation numbers for the system propagated with the optimized pulse. The ground state occupation corresponds to the (\full) line and the excited state to the (\dashed) line. (c) Convergence of $J_1$ [(\full) line], $J$ [(\dashed) line], and the fluence $E_0$ [(\dotted)~line]. (d) $J_1$ (\opencircle) and $E_0$ (\opensquare) for different penalty factors after 2000 iterations. The filled symbols [$J_1$: (\fullcircle), $E_0$: (\fullsquare)] correspond to 100 iterations.}
\label{fig:2levopt}
\end{figure}

Our goal is to find a laser pulse which transfers the ground state to the excited state $|b\rangle$ at  $T=400$. For this purpose, we use the algorithm described in equation \eref{eq:schemeZBR98} with the penalty factor $\alpha=1.0$ and the initial guess $\epsilon(t)=0.05$. After $5000$ iterations we obtain an excited state occupation of $0.9996$ and the laser field shown in \fref{fig:2levopt1}. Note that the functional tries to find a laser field which produces a high occupation and has a low fluence. This behavior is clearly visible in \fref{fig:2levopt3}, where the target yield $J_1 = | \langle b|\Psi(T) \rangle|^2$ [(\full)~line] jumps to $0.9945$ after the first iteration which corresponds to a fluence of $E_0=0.9204$ [(\dotted) line], while the converged field has the fluence of $E_0=0.0786$.  The optimal laser field [see \fref{fig:2levopt1}] has a constant amplitude of $A=0.02$ and frequency $\omega=0.1568$
\bea
\nonumber
J=\left|\langle b|\Psi(T) \rangle\right|^2 - \alpha \int_0^T \!\! dt \,\, \epsilon^2(t), 
\eea
where we have dropped the third term $J_3$ since it is always zero.

The optimal pulse within the validity of the rotating-wave-approximation (RWA) has to have a constant envelope $A(t)=const$. This can be understood with the help of the pulse-area theorem \cite{HJ94},
\bea
\label{eq:areatheorem}
\mu \int_0^T \!\! dt \,\,  A(t) = \pi ,
\eea
which states that the inversion of a two-level system (within the RWA) is achieved if the area under the pulse envelope $A(t)$ multiplied with $\mu$, i.e., the dipole matrix element, becomes $\pi$. Now consider the following functional:
\bea
\nonumber
L = \int_0^T \!\! dt \,\, A^2(t)  - \lambda  \left(  \mu  \int_0^T \!\! dt \,\, A(t) - \pi \right).
\eea
The variation of $L$ with respect to $\lambda$ yields the pulse-area theorem [equation \eref{eq:areatheorem}], while the variation with respect to the pulse shape $A(t)$ results in
\bea
\label{eq:globalopt1}
2 A(t) = \lambda \mu. 
\eea
If we plug this result back into equation \eref{eq:areatheorem} and solve for $\lambda$ we find
\bea
\label{eq:globalopt2}
\lambda &=& \frac{2 \pi}{\mu^2 T} \\
\label{eq:globalopt3}
&\Rightarrow& A(t) = \frac{\pi}{\mu T}.
\eea 
We may argue that the numerical algorithm has converged to the optimal pulse, because the RWA is perfectly valid for the chosen pulse length. 

We have run the optimization for different values $\alpha$ of the penalty factor. The occupation $J_1$ (\fullcircle) and the fluence (\fullsquare) after $100$  and $2000$  iterations (\opencircle,\,\opensquare) are shown in \fref{fig:2levoptscan}. We observe that $100$ iterations are not enough to maximize the functional for small penalty factors. Although the occupation jumps to values $J_1 >0.99$ within a few iterations, the comparison with the longer iteration shows that there is still room for improvement. Iterating long enough yields a similar fluence for a range of $\alpha$ from $0.8$ to $6.0$. 
For penalties $\alpha > 2.0$ the occupation starts to drop significantly because the fluence term is over-weighted. 
Selecting to small penalty factors leads to numerical instabilities which can be compensated by increasing the numerical accuracy of the propagation algorithm until the propagation becomes to costly.
 
The applied optimization method does not provide a possibility to find optimal fields with a predefined fluence $E_0$. This can be achieved only indirectly by the penalty factor. However, the mapping between the penalty factor $\alpha$ need not to be invertible. A practicable way to find laser pulses with a given fluence is presented in section \sref{sec:ffluence}.  

\subsubsection{Short time transfer}
In the weak field regime, i.e., where the propagation time is usually long enough to justify the use of the RWA, the application of OCT for two-level systems does not seem to be appropriate due to the large numerical effort compared to the simple formula [equation \eref{eq:areatheorem}].

 In \tref{tab:OCTvsRWA} we show a comparison between the target state occupation achieved with pulses obtained from equation \eref{eq:areatheorem} and from OCT. 
The results show that for high requirements on the inversion efficiency ($>0.995$) OCT becomes inevitable already for five-cycle pulses ($T=200$). If one requires an inversion efficiency around $0.90$ the RWA is appropriate up to single-cycle pulses ($T=40$). 

The superiority of the OCT method for short pulses will become even more apparent for a system with more than two levels \cite{WG2005}. In this case a high strength of the field (oscillating with resonance frequency) will result in excitations to other levels which are minimized by the OCT pulses.

\begin{table}[h]
\begin{indented} 
\item[]
\caption[Optimal control versus pulse-area theorem.]{\label{tab:OCTvsRWA}Comparison of the yield $P=|\langle b | \Psi(T)\rangle |^2$ when propagated with the laser field obtained with the two-level (RWA) estimate versus the laser field from OCT. Note that the period of the oscillation with $\omega_{ba}$ is $T_p=40.08$. In all optimal control runs we set the number of iterations to $5000$. In columns four and five we show the fluences calculated from the RWA and the optimized pulse, respectively. The penalty factor (in the last column) was chosen to give the best occupation for each optimization and a fluence comparable to the RWA values.}
\lineup
\begin{tabular}{@{}llllll}
\br
 $T$   &  $P_{\mathrm{RWA}}$  & $P_{\mathrm{OCT}}$ & $E_{0_{\mathrm{RWA}}}$ & $E_{0_{\mathrm{OCT}}}$ & penalty  \\
\mr
  $400$ &  $0.9986$   & $0.9996$  & $0.0803$ & $0.0786$ & $1.0$   \\    
  $200$ &  $0.9944$   & $0.9996$  & $0.1606$ & $0.1592$ & $0.5$   \\
  $100$ &  $0.9774$   & $0.9991$  & $0.3212$ & $0.3402$ & $0.3$   \\
  $50$  &  $0.9897$   & $0.9996$  & $0.6417$ & $0.7743$ & $0.3$   \\
  $40$  &  $0.8567$   & $0.9917$  & $0.8030$ & $0.7091$ & $0.3$   \\
  $25$  &  $0.7696$   & $0.9990$  & $1.2430$ & $1.4569$ & $0.3$  \\
\br
\end{tabular}
\end{indented}
\end{table}
%
\subsection{Example: Asymmetric double well}
\label{sec:adw_example}
\begin{figure}[!h]
\centering
\includegraphics*[width=.43\textwidth]{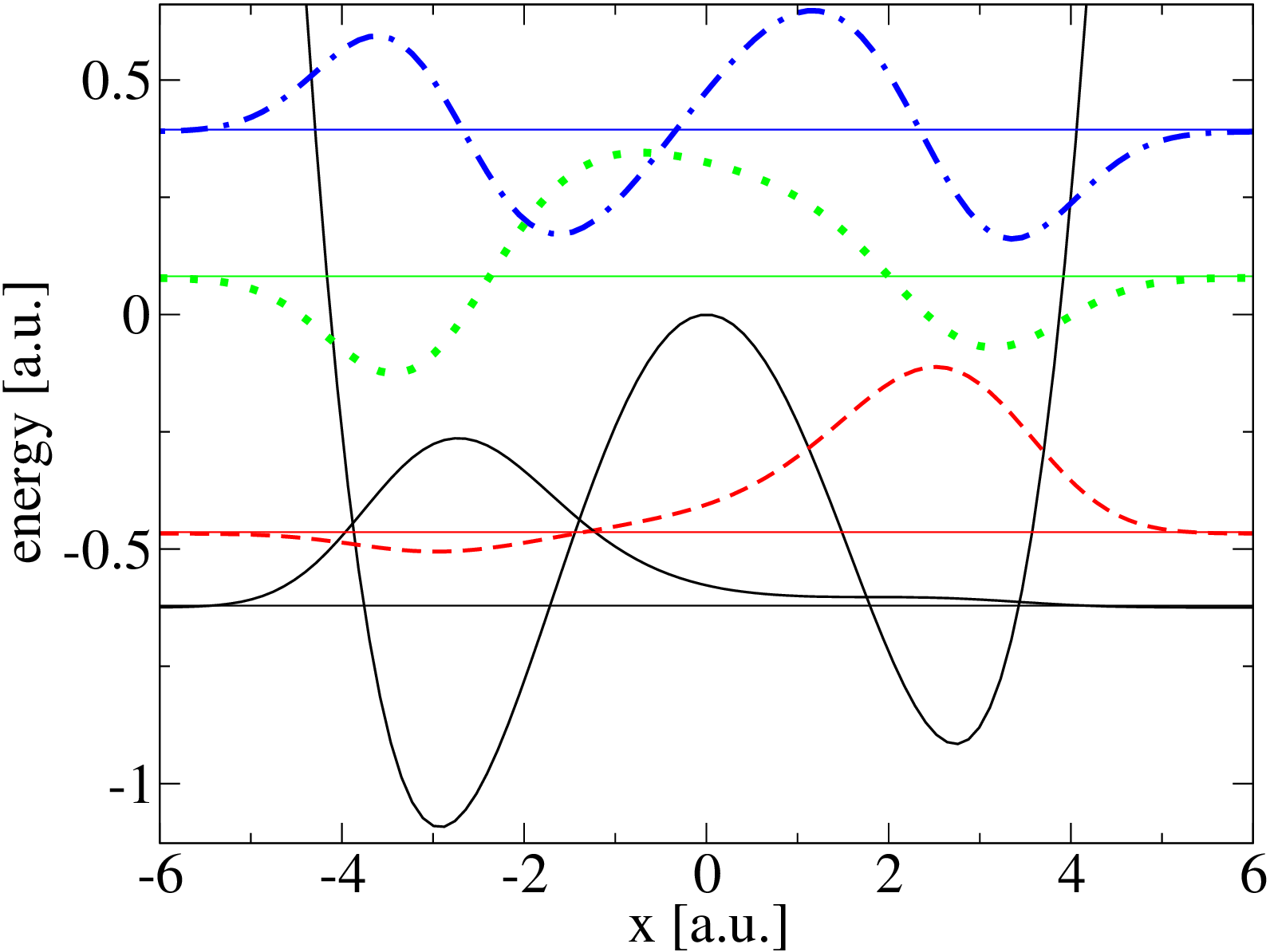}       

\caption{The plot shows the model potential with the ground state (\full), the first excited state (\dashed), the second excited state (\dotted) and the third excited state (\chain). Each state is shifted according to its eigenvalue.} 
\label{fig:potential_states}
\end{figure}

%
In the remaining examples we will focus on a one-dimensional asymmetric double well to test our algorithms. The double well is similar to that in reference \cite{GH98} but features an additional cubic term:
\begin{eqnarray}
V(x) = \frac{w_0^4}{64 B} x^4 -  \frac{\omega_0^2}{4} x^2 + \beta x^3 ,
\end{eqnarray} 
with $\omega_0$ corresponding to the classical frequency at the bottom of the well and the parameter $B$ adjusting the barrier height. The number of pairs of states below the barrier is approximately $B/\omega_0$. Here, we choose $B=\omega_0=1.0$ and $\beta=1/256$ which leads to two states below the barrier, as shown in \fref{fig:potential_states}.
%
In order to analyze the laser pulses from the optimization runs we calculate the excitation energies (see \tref{tab:xenergies}) and dipole moments (see \tref{tab:dipole_moments}) of the system by propagating in imaginary time.

\begin{table}[h]
\begin{indented}
\item[]
\caption{\label{tab:xenergies}Excitation energies in atomic units~[a.u.] for the 1D asymmetric double well, calculated by imaginary time propagation.}
\lineup
\begin{tabular}{@{}lllll}
\br  
 & $|0\rangle$        & $|1\rangle$         & $|2\rangle$        & $|3\rangle$      \\
\mr
 $|0\rangle$   & 0.       &           &          &         \\
 $|1\rangle$   & 0.1568  &  0.       &          &         \\
 $|2\rangle$   & 0.7022  &  0.5454  & 0.       &         \\
 $|3\rangle$   & 1.0147  &  0.8580  & 0.3125  & 0.      \\
 $|4\rangle$   & 1.5294  &  1.3726  & 0.8273  & 0.5147 \\
\br
\end{tabular}
\end{indented}
\end{table}
%
\begin{table}[h]
\begin{indented}
\item[]
\caption{\label{tab:dipole_moments}Dipole matrix elements for the 1D asymmetric double well, calculated by imaginary time propagation.}
\begin{tabular}{@{}ld{4}d{4}d{4}d{4}d{4}}
\br
    &  |0\rangle       & |1\rangle        & |2\rangle        &  |3\rangle       &  |4\rangle  \\
\mr
  $|0\rangle$   & -2.5676  &         &         &         &        \\
  $|1\rangle$   &  0.3921  &  2.3242 &         &         &        \\
  $|2\rangle$   &  0.6382  & -0.7037 & -0.5988 &         &        \\
  $|3\rangle$   & -0.3865  & -0.4630 &  1.7051 &  0.1958 &        \\
  $|4\rangle$   & -0.1414  &  0.2118 &  0.1593 & -1.7862 & -0.0939 \\
\br
\end{tabular}
\end{indented}
\end{table}
%
%

The time-dependent Schr\"odinger equation for the 1D double well is solved on an equidistant grid, where the infinitesimal time-evolution operator is approximated by the 2nd-order split-operator (SPO) technique \cite{FMF76}:
\begin{eqnarray} 
\nonumber
\widehat{U}_{t}^{t+\Delta t}&=&\mathcal{T}\exp\left(-i \int_{t}^{t+\Delta t} \!\! dt' \,\, \widehat{H}(t')\right)\\
\nonumber
\label{eq:spo2nd}
    & \approx  & \exp(-\frac{i}{2}\, \hat {T}\,\Delta t) \exp(-i\, \hat {V}(t)\,\Delta t)
    \exp(-\frac{i}{2}\, \hat {T}\,\Delta t) + O(\Delta t^3).
\end{eqnarray}
Following the rapidly convergent scheme described in \sref{sec:malgorithm}, one needs four propagations per iteration (if we want to avoid storing the wave function). Within the 2nd order split-operator scheme each time step requires 4 Fast Fourier Transforms (FFT) \cite{FFTW98} for the backward propagations, because we have to know the wave-function and the Lagrange multiplier in real space at every time-step to be able to evaluate the field from equations \eref{eq:fieldtildeZBR98} and \eref{eq:fieldnextZBR98}. This sums up to $16$ FFTs per time step and iteration. 
\begin{table}[h]
\begin{indented}
\item[]
\caption{\label{tab:parameter}Employed numerical parameters (atomic units).}
\begin{tabular}{lll}
\br
parameter          &        &  \\
\mr
$T$                & $400.0$   &  pulse length  \\   
$x_{\mathrm{max}}$  & $30$ & grid size \\
$dx$              & $0.1172$  & grid spacing \\
$dt$              & $0.001$   & time step \\
$\epsilon^{(0)}$   & $-0.2$  & initial guess \\
$\Delta J^{k-1,k}$ & $10^{-5}$ & convergence threshold \\
\br
\end{tabular}
\end{indented}
\end{table}
%
%

The parameters used in the runs are summarized in \tref{tab:parameter}. The initial guess for the laser field was $\epsilon^{(0)}(t)=-0.2$. This choice is arbitrary but has the advantage of producing a significant occupation in the target state at the end of the pulse, necessary to get the iteration working. Although the simple choice $\epsilon^{(0)}(t)=0.0$ will work as well in most cases, it represents a minimum of the functional since initial and target state are orthonormal. Therefore the algorithm could get stuck in principle.

In the following we apply the rapidly convergent algorithm to find the optimal field that transfers the ground-state to the 1st-excited state.
Choosing the penalty factor $\alpha=2.2$ the algorithm converges after $515$ iterations to $J=0.8470$. We consider the value as converged if the change of the functional between two subsequent iterations is smaller than $\Delta J^{k-1,k} = 10^{-5}$. 

\begin{figure}[!h]
\subfigure[]{
  \label{fig:field_E1}
  \resizebox{0.48\textwidth}{.25\textheight}{\includegraphics{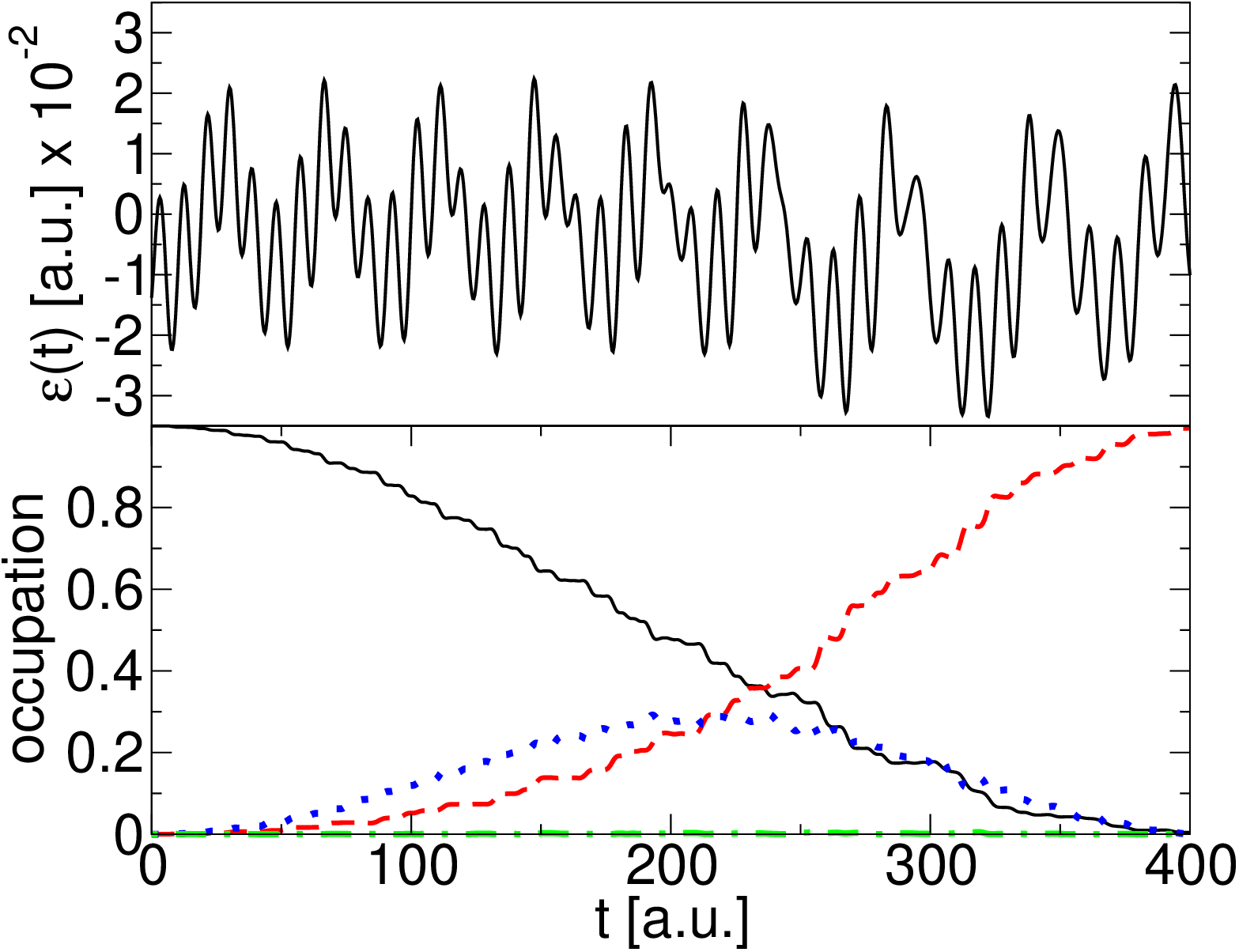}}
    }
\subfigure[]{
  \label{fig:field_E2}
  \resizebox{0.48\textwidth}{.25\textheight}{\includegraphics{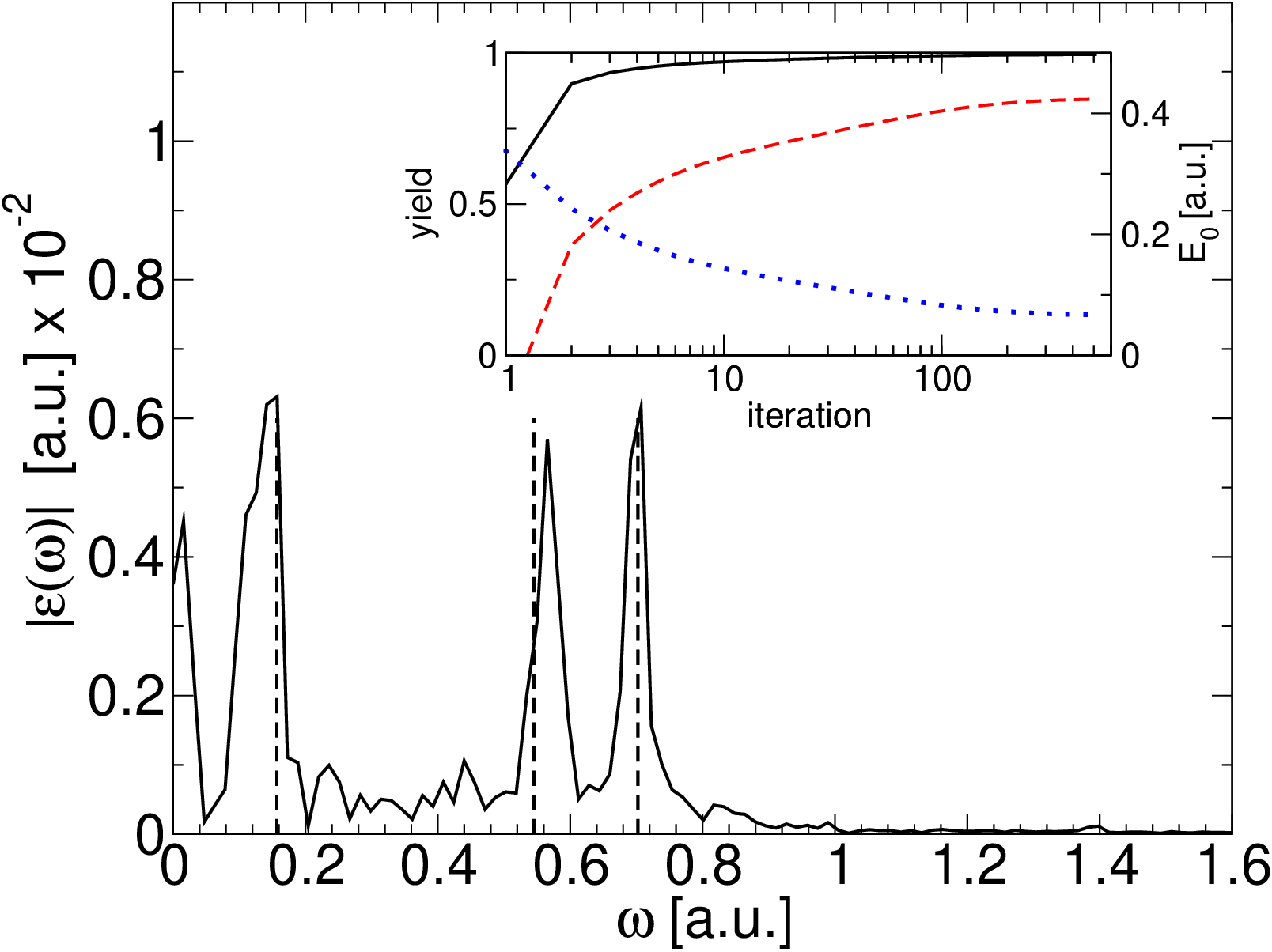}}
}
\caption[Projection operator: Optimization with standard algorithm.]{(Color online). Optimization of the $|0\ra \rightarrow |1\ra$ transition. (a) Top: Optimized field. Bottom: Time evolution of the occupation numbers $|\langle \Psi(t) | n \rangle |^2$ [$n=0$ (\full), $n=1$ (\dashed), $n=2$ (\dotted), and $n=3$ (\chain)]. (b) Spectrum of optimized field. The vertical lines indicate the transition frequencies $\omega_{01}$, $\omega_{12}$, and $\omega_{02}$. Inset: Convergence of $J_1$ [(\full) line], functional $J$ [(\dashed) line], and the fluence $E_0$ [(\dotted) line; scale on the right].}
\label{fig:field_E}
\end{figure}

 The optimal laser field is shown in the upper panel of \fref{fig:field_E1}. Applying this laser field to the system yields an occupation of $0.9944$ in the target state. The laser field exhibits a fluence of $0.0670$ which is $16\%$ less than a monochromatic pulse with a similar final occupation would need. The first step to analyze the optimal pulse is via its spectrum shown in \fref{fig:field_E2}. It is dominated by three narrow peaks which are located close to the field-free excitation energies $\omega_{01}=0.1568$, $\omega_{12}=0.5454$, and $\omega_{02}=0.7022$. Further we observe a group of peaks around $\omega_{01}$ 
($\omega\in [0,0.1]$ and $\omega \in [0.2,0.3]$) 
which do not correspond to any excitation energy of the field-free Hamiltonian. However, these frequencies play an important role in the transition process. If we filter out these frequency components, rescale the fluence to $E_0=0.0670$, and then propagate this modified laser pulse, we find at the end of the pulse the following occupations: ground-state $17.0\%$, first excited state $24.4\%$, second excited state $58.4\%$, and in all higher levels $0.2\%$. In particular, the direct transition and the back transfer from the intermediate level $|2 \rangle $ to the target state in the indirect process are less efficient without these extra frequencies. Further analysis of this kind shows that the low-frequency components and especially the zero frequency component (bias) are crucial since they introduce a (slight) shift compared to the (field-free) resonance frequencies, visible as a broadening of the $\omega_{01}$ peak in \fref{fig:field_E2}. On the other hand, if these components are missing the remaining frequencies become (in that sense) off-resonant, resulting in a low occupation of $24.0\%$ of the first excited state.   

If we filter out everything except for the extra peaks we find a target state occupation of $5\%$. Understanding these extra peaks as a third type of transfer process suggests that a mixing of transition processes in this case seems to be superior in terms of the maximum target yield per fluence than a simple monochromatic pulse.

The gain in the occupation of $0.01\%$ and in the fluence of $16.25\%$ compared to a simple monochromatic pulse has a high price -- the optimized pulse is much more difficult to realize in an experiment. Although the gain improves with shorter pulse lengths (see \tref{tab:OCTvsRWA}), this example demonstrates the typical dilemma between theory and experiment: Calculated pulses often have a far too complicated spectrum to be produced in practice. In sections \ref{sec:direct} - \ref{sec:indirect031} we demonstrate how this dilemma can be solved.

To conclude the analysis we look at the convergence behavior of the applied scheme [see inset of \fref{fig:field_E2}]. We find a fast convergence within the first 20 iterations. After these iterations the improvement of the yield slows down.

\section{Constraints on the optimal laser field}
\label{sec:ftheory}
%

%
Despite its importance only a few attempts have been made to take further restrictions on the optimal field into account. In Ref.\ \cite{K89} a scheme to calculate the pulse for a given fluence is shown. However, it does not make use of the immediate feedback introduced in Ref.\ \cite{ZBR98}, and it suffers from a rather unstable convergence. A constraint on the spectrum is considered in Ref.\ \cite{GNR92} for a steepest descent method which, in the quantum control context, also exhibits a poor convergence and a strong dependence on the initial pulse \cite{SST98}. An elegant way to restrict the spectrum has been presented in Ref.\ \cite{HMV2001}. This scheme preserves the rapid and monotonic convergence behavior of the underlying scheme \cite{ZBR98} by projecting out those parts of the time-dependent wave function which are responsible for the unwanted spectral components. However, this method is not sufficiently general and does not easily allow for an additional fluence constraint (as it keeps the penalty factor). 

The schemes shown in the following are similar to what we have discussed in Ref.~\cite{WG2005}, but are presented in a slightly more general way.
This allows us to incorporate a large variety of experimental constraints  in the optimization, for example, fluence and/or spectral constraints and phase-only shaping. As we will see in \sref{sec:direct} and \sref{sec:indirect031}, they show very good convergence, although a proof of monotonic convergence similar to reference \cite{ZR98} is not possible here. 
 The difficulty consists in $\alpha_j^{(k)}$ which is changing during the iteration and in the case of spectral constraints due to the (brute-force) modification of the field. 
But, even for the ``brute-force'' spectral filter we find a good convergence unless not too many essential features of the pulse are suppressed. 

Since we do not expect a monotonic convergence we have to add some additional intelligence to the algorithm, i.e., we store the field which produces the pulse with the highest yield in the memory. This field is considered as the result of the optimization.

\subsection{Fluence constraint}
\label{sec:ffluence}

In order to fix the fluence of the optimized laser pulse to a given value $E_0$, we have to replace the functional $J_2$ by 
\begin{eqnarray}
\label{eq:J2tilde}
\tilde{J}_2[{ \bepsilon }] &=& - \sum_{j=x,y}  \alpha_j \left[ \int_0^T \!\! dt \,\, {  \epsilon_j }^2(t) -E_{0_j} \right]. 
\end{eqnarray}
Here $\alpha_j$ is a (time-independent) Lagrange multiplier. Instead of specifying $\alpha_j$, we have to prescribe specific values $E_{0_j}$ for the components $E_{0_x}$ and $E_{0_y}$ of the laser fluence. 

Since (here) $\alpha_j$ is a Lagrange multiplier we have to vary with respect to it when calculating the total variation of $J$ [cf. equation \eref{eq:totalvarJ}]. 
The variation with respect to $\alpha_k$ results in an additional equation
\bea
\label{eq:ffvaralpha}
 \int_0^T \!\! dt \,\, {  \epsilon_k }^2(t) = E_{0_k}, \qquad k=x,y.
\eea
In the case where $\alpha_k$ is a penalty factor its value has to be set externally, while here the additional equation  can be rewritten \cite{K89}  to determine the value of the Lagrange multiplier $\alpha_k$. Inserting equation \eref{eq:ctrleps} into equation \eref{eq:ffvaralpha} yields
\bea
\nonumber
 &&\frac{1}{\alpha_k^2} \int_0^T \!\! dt \,\, \left[\underbrace{\im\langle\chi(t)|\hat{\mu}_k|\Psi(t)\rangle}_{=W_k(t)}\right]^2 = E_{0_k} \\
\label{eq:ffctrlalpha}
\Rightarrow && \alpha_k = \sqrt{ \frac{\int_0^T \!\! dt \, W_k^2(t)}{E_{0_k}}}, \qquad k=x,y,
\eea
where $\hat{\mu}_k$ is the dipole moment operator.

The remaining part of the functional stays the same, so the variations do not change, and we keep the control equations: (\ref{eq:ctrleps}),  (\ref{eq:SE}), (\ref{eq:SE2}), and (\ref{eq:SE3}).

\subsubsection{Algorithm}
\label{sec:ffalgorithm}

 The set of coupled equations which have to be solved is now given by the equations \eref{eq:ctrleps},  \eref{eq:SE}, \eref{eq:SE2}, \eref{eq:SE3}, and \eref{eq:ffctrlalpha}.
The scheme below shows the order in which these equations are solved in the $k$th step.
\begin{equation}
\label{eq:fscheme}
\begin{array}{l c c l c c c c }
  {\mbox{step k:}} \,\, & \Psi^{(k)}(0) & \overset{{\bepsilon}^{(k)}(t)}{\longrightarrow} &
 \Psi^{(k)}(T) &  &  &  &\\
          & & & \left[ \Psi^{(k)}(T) \right. & \overset{{\bepsilon}^{(k)}(t)}{\longrightarrow} &
 \left. \Psi^{(k)}(0) \right] & & \\
                   & & & \chi^{(k)}(T) & \overset{\tilde{{\bepsilon}}^{(k)}(t) 
}{\longrightarrow} & \chi^{(k)}(0), & &
\end{array}
\end{equation}
with the laser fields $\bepsilon^{(k)}(t),\tilde{\bepsilon}^{(k)}(t)$ given by
\begin{eqnarray}
\label{eq:ffeld1}
\tilde{\epsilon}_j^{(k)}(t) &=& -  \frac{1}{\alpha_j^{(k)}}\im\langle\chi^{(k)}(t)|\hat{\mu}_j|\Psi^{(k)}(t)\rangle,\\
\label{eq:ffeld2}
\epsilon_j^{(k+1)}(t) &=& \frac{\alpha_j^{(k)} }{\alpha_j^{(k+1)}} \tilde{\epsilon}_j^{(k)}(t), \qquad \qquad j=x,y,
\end{eqnarray}
where the Lagrange multiplier $\alpha^{(k+1)}_j$ is defined by
\begin{eqnarray}
\label{eq:falpha1}
\alpha_j^{(k+1)} = \sqrt{\frac{\int_0^T \!\! dt  \left[ \alpha_j^{(k)} \tilde{\epsilon}_j^{(k)}(t) \right]^2 }{E_{0_j}}}, \qquad j=x,y.
\end{eqnarray}
The initial conditions in every iteration step are
\begin{eqnarray}
\nonumber
\label{eq:finitial_psi}
  \Psi({\bf r},0)&=& \phi_i({\bf r}),\\
\label{eq:finitial_chi}
  \chi({\bf r},T) &=& \hat{O} \Psi({\bf r},T).
\end{eqnarray}
The scheme starts with the propagation of $\Psi^{(0)}(t)$ forward in time using the laser field $\epsilon^{(0)}(t)$ which has to be guessed. The result of the propagation is the wave function $\Psi^{(0)}(T)$ which is now used to calculate $\chi^{(0)}(T)$ by applying the target operator [equation \eref{eq:finitial_chi}]. We continue by propagating  $\chi^{(0)}(t)$ backwards in time using the laser field $\tilde{\epsilon}^{(0)}(t)$ defined by equation \eref{eq:ffeld1}. To solve equation \eref{eq:ffeld1}, we have to know both wave functions $\Psi^{(0)}(t)$ and $\chi^{(0)}(t)$ at the same time $t$, which makes it necessary to either store the whole time-dependent wave function $\Psi^{(0)}(t)$ or propagate it backwards with the previous laser field $\epsilon^{(0)}(t)$. The avoided storage is indicated by the brackets in the scheme (\ref{eq:fscheme}). Moreover, it is necessary to provide an initial value for $\alpha_j^{(0)}$ which we choose to be
\begin{eqnarray}
\nonumber
\alpha_j^{(0)} = \sqrt{\frac{\int_0^T \!\! dt \,\,  \left[\epsilon_j^{(0)}(t)\right]^2}{E_{0_j}}}, \qquad j=x,y.
\end{eqnarray}
The result of the backward propagation $\chi^{(0)}(t)$ is the laser field $\tilde{\epsilon}^{(0)}(t)$ which is rescaled to the right value with equation \eref{eq:ffeld2} giving $\epsilon^{(1)}(t)$. This completes the first iteration step. The second $(k=1)$ or, in general, the $k$th step repeat the described procedure starting with the initial state $\Psi^{(k)}(0)=\phi_i$ and the rescaled field $\epsilon^{(k)}(t)$.

The scheme described above has some aspects in common with the techniques described in Refs.\ \cite{K89,ZBR98}.
The basic idea of incorporating fluence constraints in the optimization algorithm was given in Ref.\ \cite{K89}. However, in contrast to this reference we make use of immediate feedback [see equation \eref{eq:ffeld1}], i.e., the backward propagation is accomplished by updating $\chi(t)$ and $\epsilon(t)$ in a self-consistent way, which was suggested in Ref.\ \cite{ZBR98}. On the other hand, the technique presented by the authors of Ref.\ \cite{ZBR98} does not allow to build in fluence constraints, since $\alpha_j$ is not a Lagrange multiplier in their case.
Roughly speaking, the technique presented above is a combination of these approaches. 
%

\subsection{Generalized filtering technique}
\label{sec:ffilter}
%
In contrast to the fluence constraint, the following general technique is not derived from a functional. Rather, a general filter is applied ``brute force'' in every iteration to the laser field,
\bea
\nonumber
\epsilon_{\mathrm{out,j}}(t)  = \mathcal{G}[\epsilon_{\mathrm{in,j}}(t)].
\eea
%
In principle, the filter $\mathcal{G}$ can be any operator. 
A few examples are discussed in sections \ref{sec:ffalgorithmspec}-\ref{sec:fffcomb}.

Since the functional itself stays the same as the standard functional [equations \eref{eq:j1}- \eref{eq:stdfunctional}], we have to solve the usual set of control equations: (\ref{eq:ctrleps}),  (\ref{eq:SE}), (\ref{eq:SE2}), and (\ref{eq:SE3}).

\subsubsection{Algorithm}
\label{sec:ffalgorithm2}
The algorithm with built in general filtering is similar to the one presented in the previous section (\sref{sec:ffluence}) with two important differences: The factor $\alpha_j$ is a penalty factor. It has to be specified from the start and remains unchanged during the optimization. Second, the update of $\epsilon_j^{(k+1)}(t)$ in equation \eref{eq:ffeld2} is replaced by
\begin{eqnarray}
\label{eq:gfilter}
\epsilon_j^{(k+1)}(t) &=&  \mathcal{G} \left[ \tilde{\epsilon}_j^{(k)}(t)\right],  \qquad j=x,y,
\end{eqnarray}
where the symbol $\mathcal{G}$ indicates a given filter operator acting on the field $\tilde{\epsilon}_j^{(k)}(t)$.
%
\subsubsection{Spectral constraints}
\label{sec:ffalgorithmspec}
%
If spectral filtering is required, we formulate the constraint with the help of a filter function $f_j(\omega)$, the Fourier transform $\mathcal{F}$, and its inverse $\mathcal{F}^{-1}$. Equation \eref{eq:gfilter} is now replaced by
\begin{eqnarray}
\label{eq:feld2_filter}
\epsilon_j^{(k+1)}(t) &=&  \mathcal{F}^{-1} \left[ f_j(\omega) \, \mathcal{F}\left[\tilde{\epsilon}_j^{(k)}(t)\right]\right],  \qquad j=x,y.
\end{eqnarray}
Since $\epsilon_j(t)$ is real-valued we have to make sure that $f_j(\omega) = f_j(-\omega)$. For example, the filter function could be chosen to be 
\begin{eqnarray}
\label{eq:filter1}
f_j(\omega) =  \exp[-\gamma (\omega-\omega_0)^2] + \exp[-\gamma (\omega+\omega_0)^2],
\end{eqnarray}
so that only the components around the center frequency $\pm \omega_0$ are allowed in the pulse. If one uses instead
\begin{eqnarray}
\label{eq:filter2}
\tilde{f}_j(\omega) = 1- \left(\exp[-\gamma (\omega-\omega_0)^2] + \exp[-\gamma (\omega+\omega_0)^2]\right),
\end{eqnarray}
one would allow every spectral component in the laser field except the components around $\pm \omega_0$.\\
In Ref.~\cite{janphd} we show that the spectral filter technique can be derived from a modified form of the standard functional. 
%
\subsubsection{Laser-envelope constraints}
\label{sec:ffftd}
Even though the formulation of the theory with a time-dependent penalty factor provides already one way to enforce a time-dependent shape function $h_j(t)$, we want to present an alternative way. Here, we replace equation \eref{eq:gfilter} by
\bea
\label{eq:tdenvelope}
\epsilon_j^{(k+1)}(t) &=&  h_j(t) \, \tilde{\epsilon}_j^{(k)}(t), \qquad j=x,y.
\eea
%
While the more elegant way is to use a time-dependent penalty factor \cite{SV99}, its application is not always possible, for example, in the case of a fixed laser fluence where the penalty factor does not exist. While the alternative method still enables us to impose restrictions on the laser envelope and at the same time to fix the fluence to a given value.

\subsubsection{Phase-only shaping}
\label{sec:fffphase}
Many experiments are carried out using only phase-shaping, i.e., only the phases of the spectral components are optimized but the amplitude spectrum itself stays fixed (for details on the experiment see \sref{sec:cllexperiments}). This is done to reduce the enormous search space for the genetic algorithm and to achieve  a faster convergence. 

For the implementation of phase-only shaping into the computational optimization we have to replace equation \eref{eq:gfilter} by the following equations:
\bea
\label{eq:phaseonly1}
\tilde{\epsilon}_j^{(k)}(\omega) &=&  \mathcal{F}\left[\tilde{\epsilon}_j^{(k)}(t)\right],  \\ 
\label{eq:phaseonly2}
\epsilon_j^{(k+1)}(t) &=&  \mathcal{F}^{-1}\left[ A_j(\omega) \, \frac{\tilde{\epsilon}_j^{(k)}(\omega)}{|\tilde{\epsilon}_j^{(k)}(\omega)|} \right],  \qquad j=x,y,
\eea
where the function $A_j(\omega)$ contains the predefined amplitude spectrum which, in the experiment, corresponds to the spectrum of the laser pulse that enters the pulse shaping device.

\subsubsection{Combination of filters}
\label{sec:fffcomb}

In principle, the filters can be freely combined. Then
equation \eref{eq:gfilter} has to be extended to
\begin{eqnarray}
\label{eq:gfilterN}
\epsilon_j^{(k+1)}(t) &=&  \mathcal{G}_N \left[\mathcal{G}_{N-1}  \left[ \ldots \mathcal{G}_1  \left[ \tilde{\epsilon}_j^{(k)}(t)\right] \right] \ldots \right], \qquad j=x,y.
\end{eqnarray}
Care must be taken if the filters are conjugated, for example, a filter in the frequency domain will change the field in the time domain as well. In this case the order of the filters is important. For example, we want to restrict the spectrum and at the same time require a Gaussian shaped laser envelope. In general, it is impossible to satisfy both requirements at the same time. However, if the spectral filter function is broad enough (unlike a $\delta$ function which will always result in a constant envelope) the time filter can be satisfied in a reasonable way. Of course, if the filters are conjugated only the last filter will have the full desired effect.

\subsection{Generalized filtering technique with fluence constraint}
\label{sec:ffilterE0}
%
It is also possible to combine the fluence constraints with the filtering techniques. This combination makes it possible to implement even more realistic experimental constraints in computational pulse optimizations. 
Therefore we use the functional and the control equations discussed in \sref{sec:ffluence} and add the ``brute force'' method of \sref{sec:ffilter}. 
%

\subsubsection{Algorithm}
\label{sec:ffalgorithm3}

The algorithm is a simple combination of the schemes presented above. Basically we use the scheme (\ref{eq:fscheme}), equation \eref{eq:ffeld1}, and instead of equation \eref{eq:ffeld2} we employ
\begin{eqnarray}
\label{eq:feld2_filter_fixed}
\bar{\epsilon}_j^{(k)}(t) &=&   \mathcal{G} \left[ \tilde{\epsilon}_j^{(k)}(t)\right], \\
\label{eq:feld2_filter_fixed2}
\epsilon_j^{(k+1)}(t) &=&  \frac{\alpha_j^{(k)}}{\alpha_j^{(k+1)}} \bar{\epsilon}_j^{(k)}(t),  \qquad j=x,y,
\end{eqnarray}
where $\mathcal{G}$ is a given filter operator and $\alpha_j^{(k+1)} $ is evaluated inserting the filtered field $\bar{\epsilon}_j^{(k)}(t)$ in
\begin{eqnarray}
\label{eq:alpha_filterfixed}
\alpha_j^{(k+1)} = \sqrt{\frac{\int_0^T \!\! dt  \left[ \alpha_j^{(k)} \bar{\epsilon}_j^{(k)}(t) \right]^2 }{E_{0_j}}}, \qquad j=x,y,
\end{eqnarray}
to enforce the predefined value for $E_{0_j}$. 
Note that the total spectral power is related to the time-integrated quantity by Parseval's theorem,
\begin{eqnarray} 
\nonumber
E_{0_j} = \int_{-\infty}^{+\infty} \!\! dt \,\, \theta(t) \theta(T-t) \left[\epsilon_j(t)\right]^2 = \frac{1}{2 \pi} \int_{-\infty}^{+\infty} \!\! d\omega \,\, \left| \epsilon_j(\omega)\right|^2. 
\end{eqnarray}
In this combined form of the algorithm we first apply the filter function to the laser field (\ref{eq:feld2_filter_fixed}) and then rescale the field to yield the right value for $E_{0_j}$ [see equation \eref{eq:feld2_filter_fixed2}].
%
\subsection{Example: Direct Transition}
\label{sec:direct}

We now present the results of the algorithm with spectral constraints and a penalty factor. In the previous chapter  the transfer of the particle occured via a mixture of a direct transition and indirect transitions. This motivates the following aim:  Find a laser pulse that produces a high yield and that contains only spectral components centered around the resonance frequency $\omega_{01}$. 

In order to find an optimal pulse with $\omega_{01}$ as the center frequency, we use a Gaussian frequency filter $f(\omega)$ according to equation \eref{eq:filter1} centered around $\omega_0 = \omega_{01}$ and with $\gamma=500$.  

\begin{figure}[!h]
\centering 
\subfigure[]{
  \label{fig:field_F1}
      \resizebox{0.47\textwidth}{.25\textheight}{\includegraphics{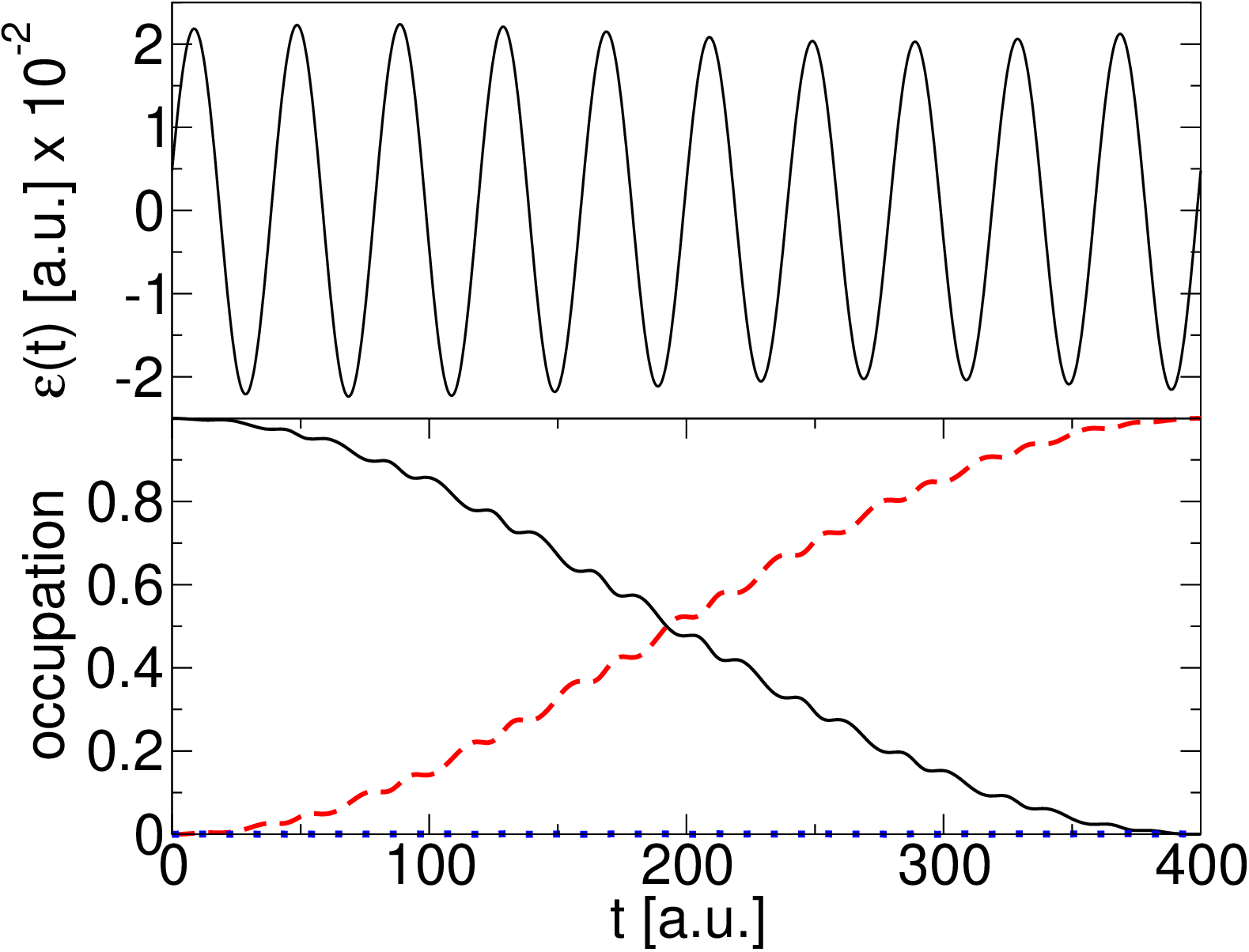}}
    }
\subfigure[]{
  \label{fig:field_F2}
  \resizebox{0.47\textwidth}{.25\textheight}{\includegraphics{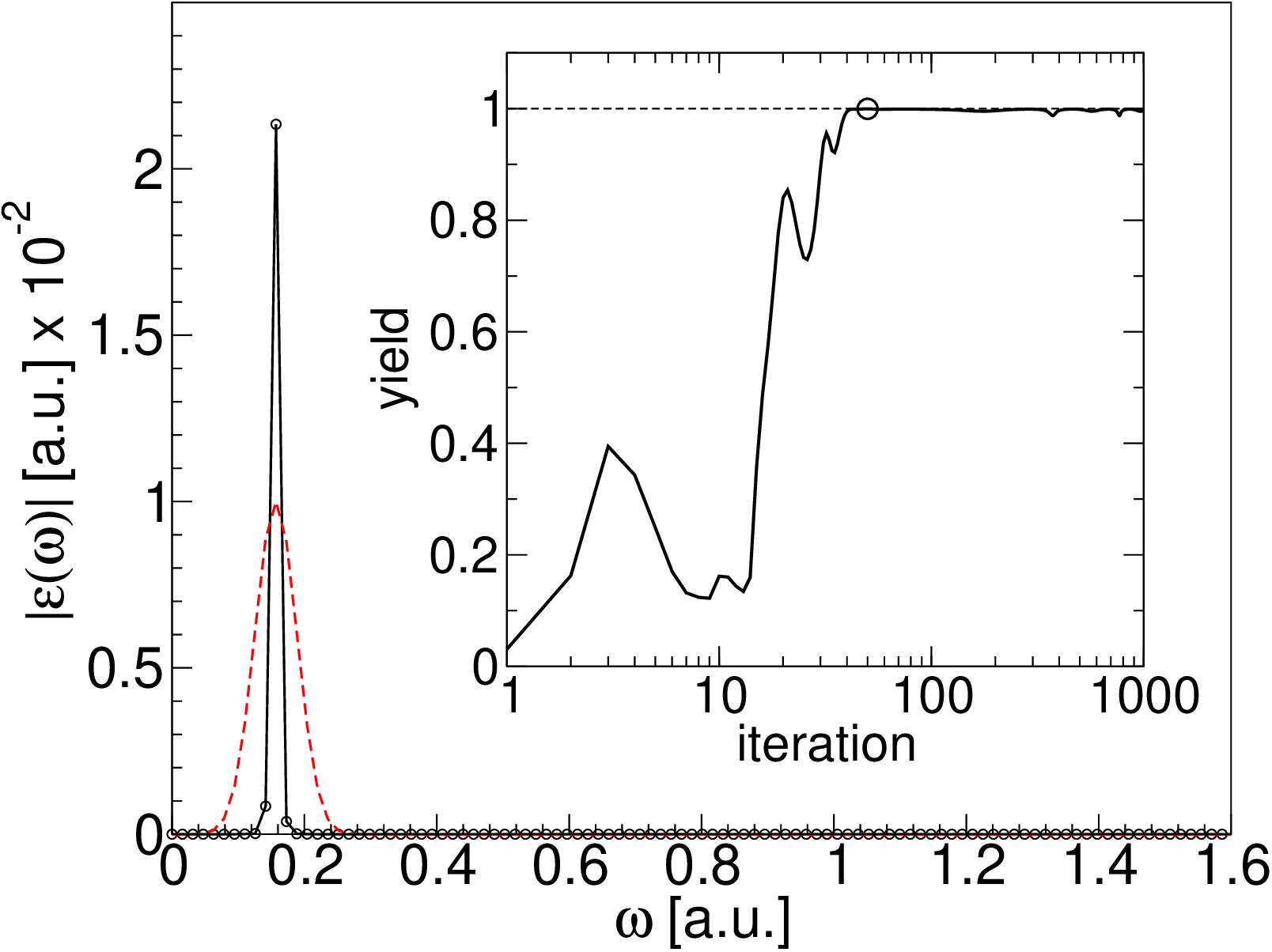}}
}
\caption[Projection operator: Filter-OCT calculation with Gaussian at resonance.]{(Color online). Optimization of the $|0\ra \rightarrow |1\ra$ transition with a Gaussian frequency filter around $\omega_{01}=0.1568$. (a) Upper panel: Optimized field. Lower panel: The time-dependent occupation numbers confirm that only the ground state (\full) and the first excited state (\dashed) take part in the transition process. The second excited state population (\dotted) is hardly visible. (c) Spectrum (\full) and filter function $f(\omega)$ (\dashed), scaled by $0.01$. Inset: Convergence of $J_1$. The (\opencircle) indicates the iteration with the highest yield.}
\label{fig:field_F}
\end{figure}

After $50$ iterations the algorithm finds a laser pulse which yields of $99.97\%$. The penalty factor has been set to $\alpha=0.05$. The obtained value for $E_0 = 0.090$ is slightly higher than the estimate from the two-level model ($E_0=0.08$, $J_1 = 99.30\%$).
The slight envelope on the field, visible in the upper panel of \fref{fig:field_F1}, is caused by the finite width of the Gaussian [see (\dashed) line in \fref{fig:field_F2}]. Frequency components near $\omega_{01}$ are still allowed in the pulse and result in a beat pattern (envelope). The time-dependent occupation numbers confirm that the higher states are not occupied during the transition [see the lower panel of \fref{fig:field_F2}]. The convergence shown in \fref{fig:field_F2} is rather smooth.
Note that if we desire a sinusoidal field with a constant envelope, we have to reduce the width of the Gaussian to a $\delta$ function to allow only one single component in the spectrum. Using such a filter we obtain a yield of $99.79\%$ and $E_0=0.085$. This field oscillates with the amplitude $A=0.0207$ which is slightly higher than the amplitude derived from the pulse area theorem [see equation \eref{eq:opt_amp_res}].

\subsection{Example: Transition via $|3\ra$}
\label{sec:indirect031}
%
\begin{figure}[!h]
\subfigure[]{
  \label{fig:field_FC031_1}
  \resizebox{0.48\textwidth}{.26\textheight}{\includegraphics{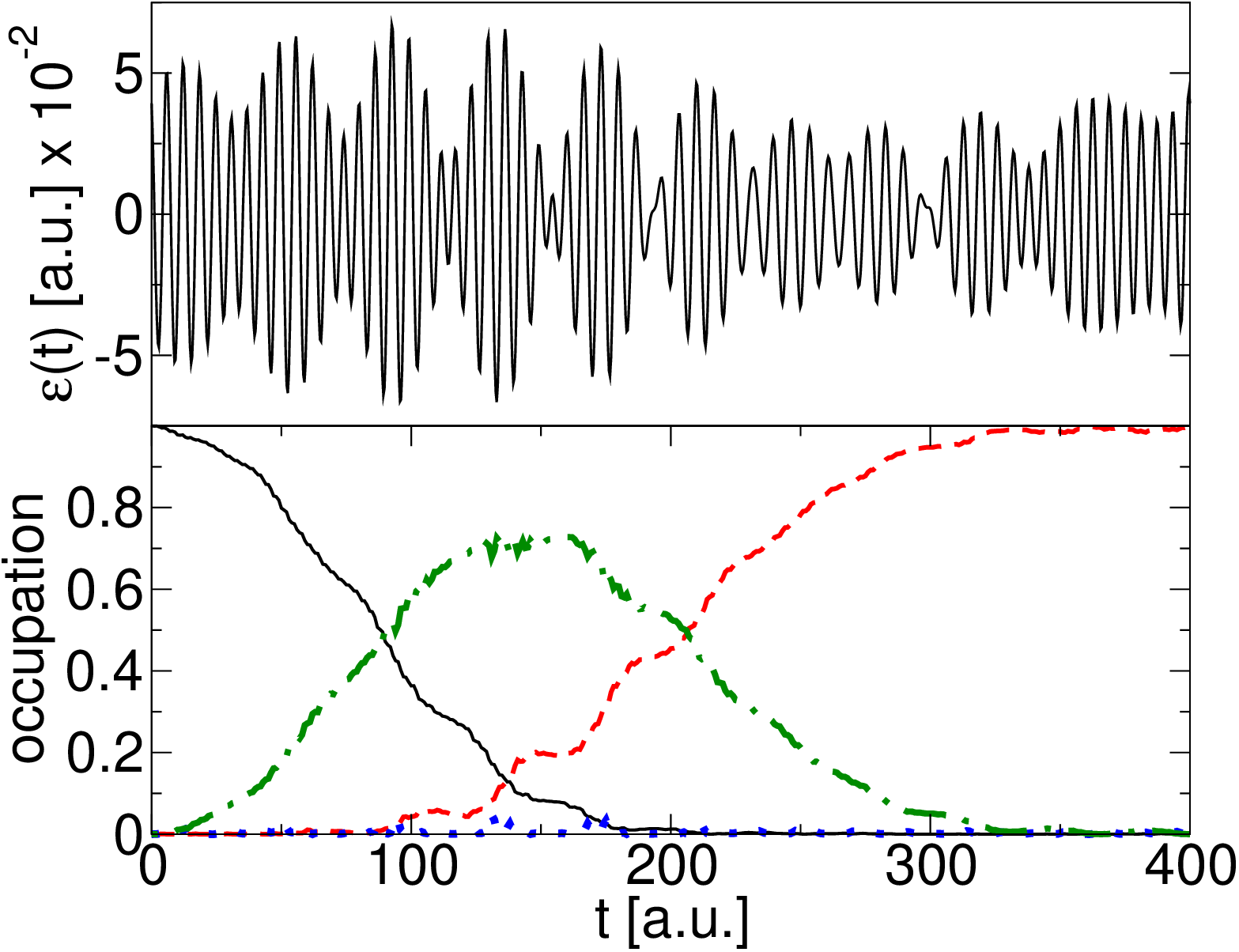}}
}
\subfigure[]{
  \label{fig:field_FC031_2}
  \resizebox{0.48\textwidth}{.26\textheight}{\includegraphics{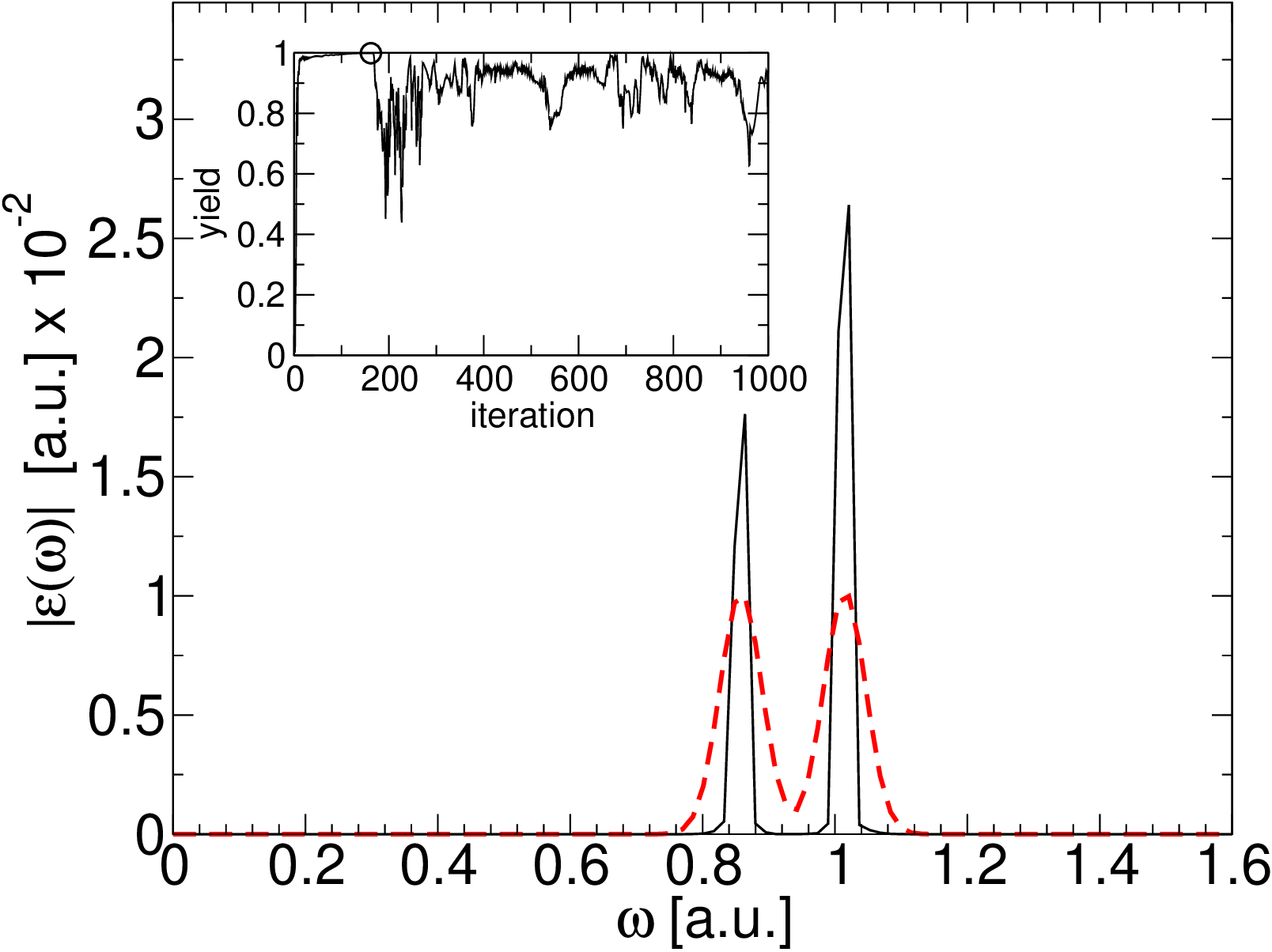}}
}
\caption[Projection operator: Optimization via intermediate state $|3\ra$.]{(Color online). We apply the optimization algorithm for the transition $|0\ra \rightarrow |1\ra$ with a double Gaussian frequency filter allowing only frequencies around $\omega_{03}$ and $\omega_{31}$, and in addition we set $E_0=0.320$. (a) Optimized field and time-dependent occupation numbers [ground state (\full), first excited state (\dashed), and second excited state (\dotted)]. (b) Spectrum and filter function $f(\omega)$ (\dashed), scaled by $0.01$. Inset: Convergence of $J_1$. The (\opencircle) indicates the iteration with the highest yield. }
\label{fig:field_FC031}
\end{figure}

The process $|0\rangle \to |3\rangle \to |1\rangle$ using the third excited state as an intermediate state plays only a minor role in the examples considered above. Now, we are going to optimize the laser pulse such that the transition occurs exclusively via this process. In addition we require $E_0=0.320$. 
This time a double Gaussian filter is centered at $\omega_{13}$ and $\omega_{03}$. The width parameter is again $\gamma=500$.
 
The results are shown in \fref{fig:field_FC031}. Like in the previous example, the high restrictions within the optimization lead to a rather erratic convergence [see inset of \fref{fig:field_FC031_2}]. The field, shown in the upper panel of \fref{fig:field_FC031_1}, corresponds to the $162$th iteration and produces a target state occupation of $99.89\%$. The time-dependent occupation numbers [see lower panel of \fref{fig:field_FC031_1}] show that the transition occurs exactly in the desired way.

\section{Time-dependent control targets}
\label{sec:theoryextent}
%
The control targets we have considered so far refer to the maximization of a quantity at the end of the laser pulse, e.g., the occupation of some excited state. This kind of optimization objective is called a {\it time-independent target}, since it leaves the dynamical path the quantum system follows towards a target state undefined. In this chapter we demonstrate that it is also possible to control this path, i.e., to find the laser pulse which leads the quantum system as close as possible along a predefined trajectory \cite{SR90}. The path could be simply a trajectory in the configuration space but it may also be a path in a more abstract sense, e.g., a trajectory in quantum number space to control {\it how} a transition takes place. 

Control targets that require a time-dependent formalism are the control of bond distances in molecules (e.g., steering the fragmentation process in time or using a laser to keep a certain bond distance), the optimization of high harmonics \cite{B2000,CC2001}, and the control of currents in time, e.g., in a molecular switch. 

To our knowledge, three different methods for the control of time-dependent targets have been proposed so far: A 4th-order Euler-Lagrange equation to determine the envelope of the control-field has been derived in Ref.\ \cite{GGB2002}. However, it is restricted to very simple quantum systems.

A very elegant method, known as tracking has been proposed in Refs.\ \cite{ZR2003,S2003}. Despite its tremendous success, this method bears an intrinsic difficulty: One has to prescribe a path that is actually achievable with a laser field, otherwise singularities in the field appear due to the one-to-one correspondence between the laser field and the given trajectory. In practice, this may require a lot of intuition. However, the most severe drawback is that tracking cannot be used together with constraints on the laser pulse (see \sref{sec:ftheory}), which is again a problem of the one-to-one correspondence. 

The third method is an optimal control scheme for time-dependent targets \cite{OTR2004,mySWG2005}. The new method is monotonically convergent and in contrast to tracking it does not rely on one's intuition in choosing the right control targets. Furthermore, the method is not restricted to two-level systems and can be extended to incorporate fluence and spectral constraints, as we will show in \sref{sec:tdtargetsconstraints}. First applications of the method can be found in \cite{OTR2004,KM2004,ioanadipl,janphd,WG2006spie}. 


In \sref{sec:theoryextent} we present the formalism for the control of time-dependent targets \cite{SR90} in the same way as in \sref{sec:stdoct}.
The algorithm \cite{OTR2004, ioanadipl} to solve the resulting equations is described in \sref{sec:tdctrlalg}. 

In  \sref{sec:tdtargetsconstraints} we present a novel algorithm which combines the standard control algorithm of \sref{sec:tdctrlalg} with the constraints discussed in \sref{sec:ftheory}. We test the developed algorithm for the asymmetric double well model (see \sref{sec:adw_example}) and discuss the results for the control of a path in quantum number space in \sref{sec:tdoct_example}. 

\subsection{Derivation of the control equations}
Consider a modified form for the first term of the standard functional in equation \eref{eq:stdfunctional}
\begin{eqnarray} \label{eq:J1tdctrl}
J_1[\Psi]&=&\frac{1}{T} \int_0^T \!\! dt \,\, w(t) \langle \Psi(t)|\hat{O}(t)|\Psi(t)\rangle,
\end{eqnarray} 
with $w(t)$ representing a time-dependent weight function which is normalized in the following way
\bea
\nonumber
\int_0^T \!\! dt \,\, w(t) = T, \qquad w(t) \geq 0 \quad \forall \,\,  t.
\eea
 The weight function $w(t)$ is supposed to steer the relative importance of the time-dependent target operator. If the target operator $\hat{O}(t)$ is positive semidefinite then $J_1$ will reach its maximum, if at each point in time the expectation value of the operator $\langle \Psi(t)|\hat{O}(t)|\Psi(t)\rangle$ is maximized. For the moment we will keep the operator as general as possible and postpone the discussion of different examples to \sref{sec:tdctrltarget}. 

The other parts of the standard functional, $J_2$ in equation \eref{eq:penalty} and $J_3$ in equation \eref{eq:j3}, remain unchanged. Thus, the functional derivative of $J$ with respect to $\Psi$ becomes
\bea
\nonumber
\frac{\delta J_1}{\delta \Psi(\brr',\tau)} &=& \frac{1}{T} \, w(\tau) \hat{O}(\tau) \Psi^*(\brr',\tau) \, ,\\
\nonumber
\frac{\delta J_3}{\delta \Psi(\brr',\tau)} &=& -\rmi \left(\rmi \partial_\tau + \hat{H}(\tau) \right) \chi^*(\brr',\tau) - \left[\chi^*(y,t) \delta(t-\tau)\right]\Big|_0^T,
\eea
which yields
\bea
\nonumber
\delta_{\Psi} J &=& \int_0^T \!\! d\tau \left\{\frac{1}{T} \, w(\tau) \left\langle \Psi(\tau)|\hat{O}| \delta \Psi(\tau) \right\rangle  + \rmi \left\langle \left( \rmi \partial_\tau - \hat{H}(\tau) \right) \chi(\tau) |\delta \Psi(\tau) \right\rangle \right\} \\
\label{eq:varpsi2}
&& \qquad - \left\langle \chi(T) | \delta \Psi(T) \right\rangle + \underbrace{\left\langle \chi(0) | \delta \Psi(0) \right\rangle}_{=0} .
\eea

\subsection{Time-dependent  control equations}
\label{sec:tdctrleq}
%
Setting the variation with respect to $\Psi$ in equation \eref{eq:varpsi2} equal to zero, we obtain an inhomogeneous \TDSE~for the Lagrange multiplier $\chi({ \bf r },t)$
\bea
\label{eq:SEchi}
\left( \rmi \partial_t - \hat{H}(t) \right) \chi({ \bf r },t) &=& - \frac{\rmi}{T} w(t) \hat{O}(t) \Psi(\brr,t)\, , \qquad 
\chi({ \bf r },T) = 0.
\eea
Its solution can be formally written as
\begin{eqnarray}
\label{eq:SOL_INHSE}
\chi({ \bf r },t)&=&\hat{U}_{t_0}^{t}\chi({ \bf r },t_0)-
       \frac{1}{T}\int_{t_0}^{t} \!\! d\tau \,\, \hat{U}_{\tau}^{t}\left(w(\tau) \hat{O}(\tau) \, \Psi({ \bf r },\tau)\right),
\end{eqnarray}
where $\hat{U}_{t_0}^{t}$ is the time-evolution operator \cite{Sakurai} defined by
\bea
\nonumber
 \hat{U}_{t_0}^{t}=\mathcal{T}\exp\left[-\rmi \int_{t_0}^{t} \!\! dt' \,\, \hat{H}(t')\right].
\eea
Since the other parts of the functional correspond to the standard functional, also the variations with respect to $\chi({ \bf r },t)$ and $\epsilon_k(t)$ are identical to equations \eref{eq:SE} and \eref{eq:ctrleps}, which we restate here for convenience
\bea
\label{eq:tdctrlSE}
\left( \rmi \partial_t - \hat{H}(t) \right) \Psi({ \bf r },t)  &=& 0, \qquad 
\Psi({ \bf r },0) = \phi_i({ \bf r }),\\
\label{eq:tdctrlfield}
 \alpha_k \epsilon_k(t) &=& -\im\,\langle\chi(t)|\hat{\mu}_k|\Psi(t)\rangle,  \qquad k=x,y. 
\eea
Similar to \sref{sec:ffluence} we can introduce fluence constraints by using  $\tilde{J}_2$ [equation \eref{eq:J2tilde}] instead of $J_2$ [equation \eref{eq:penalty}], for which we obtain an additional equation from the variation with respect to the Lagrange multiplier $\alpha_k$
\bea
\label{eq:tdctrlalpha}
\int_0^T \!\! dt \,\, {  \epsilon_k }^2(t) = E_{0_k} .
\eea 
The set of equations that we need to solve is now complete:  (\ref{eq:SEchi}), (\ref{eq:tdctrlSE}), (\ref{eq:tdctrlfield}), and in the case of a predefined  fluence, we have in addition equation \eref{eq:tdctrlalpha}.

\subsection{Target operators}
\label{sec:tdctrltarget} 
The physical meaning of the functional $J_1$ given by equation \eref{eq:J1tdctrl} depends on the choice of the target operator. In the following we present the most important choices and discuss their physical interpretation.

\paragraph{Final-time control:}
Since our approach is a generalization of the standard optimal control formulation given in \sref{sec:stdoct} we first observe that the latter is trivially recovered as a limiting case by setting
\begin{eqnarray}
\nonumber
w(t) = 2 T \delta(t-T), \qquad \hat{O} = | \phi_f \rangle \langle \phi_f|,
\end{eqnarray}
where we use the definition: $\int_0^T \!\! dt \, \delta(t-T) = 1/2$.
Here $|\phi_f \ra$ represents the target state, which the propagated wave function $\Psi(t)$ is supposed to reach at time $T$. 
In this case the target functional reduces to \cite{K89,ZBR98}
\begin{eqnarray}
\nonumber
J_1 = | \langle \Psi(T) | \phi_f \rangle| ^2 .
\end{eqnarray}
The target operator may also be local, as pointed out in  Ref.\ \cite{ZR98}. If we choose $w(t) = 2 T \delta(t-T)$ and $\hat{O} = \delta({ \bf r }-{ \bf r }_0)$ (the density operator), we can maximize the probability density in ${ \bf r }_0$ at $t=T$,
\begin{eqnarray}
\label{eq:loc_op}
J_1 =  \langle \Psi(T)| \hat{O} |\Psi(T) \rangle = n({ \bf r }_0,T). 
\end{eqnarray}
In the actual calculations, the $\delta$~function can be approximated by a narrow Gaussian.

\paragraph{Maximizing the average:}
In the literature \cite{GGB2002,OTR2004} the functional (\ref{eq:J1tdctrl}) has so far only been used with a time-{\underline{in}}dependent target operator, e.g.,
\bea
\nonumber
\hat{O} = | \phi_f \rangle \langle \phi_f|,
\eea 
combined either with a time-independent ($w(t)=1$) or time-dependent weight function \cite{KM2005}.

In the first case, we require the laser pulse to maximize the average occupation in state $|\phi_f \rangle$, i.e., the earlier the laser pulse drives the initial to the target state (and keeps it there), the higher the yield. 


\paragraph{Wave function follower:}
The most ambitious goal is to find the pulse that forces the system to follow a predefined wave function $\phi({ \bf r },t)$. If we choose
\begin{eqnarray}
\nonumber
w(t) &=& 1, \\
\hat{O}(t) &=& |\phi(t)\rangle\langle\phi(t)|, 
\end{eqnarray}
the maximization of the time-averaged expectation value of $\hat{O}(t)$ becomes almost equivalent to the inversion of the \TDSE, i.e., for a given function $\phi({ \bf r },t)$ we find the field $ { \boldsymbol \epsilon }(t)$ 
so that the propagated wave function $\Psi({ \bf r },t)$ comes as close as possible to the target $\phi({ \bf r },t)$ in the space of admissible control fields.
We can apply this method to the control of time-dependent occupation numbers, if we choose the time-dependent target to be 
\begin{eqnarray}
\label{eq:td_op_numbers}
| \phi(t) \rangle &=&c_0(t)e^{-i\mathcal{E}_0t}|0\rangle +c_1(t)e^{-i\mathcal{E}_1t}|1\rangle +c_2(t)e^{-i\mathcal{E}_2t}|2\rangle +\ldots\:\:,\\\nonumber
\hat{H}_0|n\rangle&=&\mathcal{E}_n|n\rangle\\
\label{eq:td_op}
\hat{O}(t)&=& | \phi(t) \rangle \langle \phi(t) |.
\end{eqnarray}
The functions $|c_0(t)|^2, |c_1(t)|^2, |c_2(t)|^2, \ldots$ are the predefined time-dependent level-occupations which the optimal laser pulse will try to achieve. In general the functions $c_0(t),c_1(t),c_2(t), \ldots$ can be complex, but as demonstrated in Ref.\,\cite{mySWG2005}, real functions can be sufficient to control the occupations in time. For example, if in a two-level system the occupation is supposed to oscillate with frequency $\Omega$ we could choose $c_0(t) = \cos(\Omega t)$ and, by normalization,  $c_1(t)=\sin(\Omega t)$. This defines a time-dependent target operator by equations \eref{eq:td_op_numbers} and \eref{eq:td_op}.

\paragraph{Moving density:}
The operator used in equation \eref{eq:loc_op} can be generalized to
\begin{eqnarray}
\label{eq:tdmovdens1}
\hat{O}(t)&=&\delta({ \bf r }-{\bf r }_0(t)),\\
J_1 &=& \frac{1}{T} \int_0^T \!\! dt \,\, w(t) \langle \Psi(t) |  \delta({ \bf r }-{ \bf r}_0(t)) | \Psi(t) \rangle  \nonumber\\ 
\label{eq:tdmovdens}
&=& \frac{1}{T} \int_0^T \!\! dt \,\, w(t) n({\bf r}_0(t),t).
\end{eqnarray}
Here $J_1$ is maximized if the field focusses the density at each point in time in ${\bf r }_0(t)$.  An example can be found in Ref.\ \cite{WG2006spie}. The relative importance of different time intervals can then be adjusted by the weight function $w(t)$.

\subsection{Algorithm for time-dependent control targets}
\label{sec:tdctrlalg}
Equipped with the control equations (\ref{eq:SEchi}), (\ref{eq:tdctrlSE}), (\ref{eq:tdctrlfield}) we have to find an algorithm to solve these equations for ${\boldsymbol \epsilon}(t)$.
In the following we describe such a scheme which is similar to those in Refs.\ \cite{MT2003,OTR2004} where the additional parameters $\eta$ and $\xi$ have been ``artificially'' introduced (not derived by a functional variation) to ``fine tune'' the convergence of the algorithm. A monotonic convergence in $J$ can be proven if $\eta \in [0,1]$ and $\xi \in [0,2]$ \cite{MT2003}. Here, $\boldsymbol \alpha$ is always a penalty factor.

The algorithm starts with propagating $\Psi^{(0)}(0)=\phi_i$ forward in time with an initial guess for the laser field $\bepsilon(t)^{(0)}$,
\begin{equation}
\nonumber
\begin{array}{l c c l c c c c }
  {\mbox{step 0:}} \,\, & \Psi^{(0)}(0) & \overset{{\bepsilon}^{(0)}(t)}{\longrightarrow} &
 \Psi^{(0)}(T). &  &  &  &
\end{array}
\end{equation}
The backward propagation of $\chi^{(0)}(t)$ is started from $\chi^{(0)}(T)=0$ solving an inhomogeneous \TDSE~which requires $\Psi^{(0)}(t)$ as input (with $k=0$),
\begin{equation}
\label{eq:tdctrlstepk}
\begin{array}{l c c l c c c c }
 {\mbox{step k:}} \,\, & & & \left[ \Psi^{(k)}(T) \right. & \overset{{\boldsymbol \epsilon}^{(k)}(t)}{\longrightarrow} &
 \left. \Psi^{(k)}(0) \right] & & \\
                   & & & \chi^{(k)}(T) & \overset{\widetilde{{\boldsymbol \epsilon}}^{(k)}(t)
,\:\:\Psi^{(k)}(t)\:\: }{\longrightarrow} & \chi^{(k)}(0)\, . & &
\end{array}
\end{equation}
The brackets indicate that the storage of the wave function $\Psi^{(0)}(t)$ can be avoided if we propagate it backwards in time as well using $\bepsilon^{(0)}(t)$. The backward propagation of $\chi^{(0)}(t)$ requires the laser field determined by ($k=0$),
\begin{eqnarray}
\label{feld1}
\widetilde{\epsilon}_j^{(k)}(t) &=& (1-\eta)\epsilon_j^{(k)}(t) 
-  \frac{\eta}{\alpha_j}\im\, \langle\chi^{(k)}(t)|\hat{\mu}_j|\Psi^{(k)}(t)\rangle,\,\,\ j=x,y . \,\,\,\,\,\,\,\,\,
\end{eqnarray}
The next step is to start a forward propagation of  $\Psi^{(1)}(0) = \phi_i$ ($k=0$)
\begin{equation}
\nonumber
\begin{array}{l c c l c c c c }
                   & & & & & \left[ \chi^{(k)}(0) \right. & \overset{\widetilde{{\boldsymbol \epsilon}}^{(k)}(t) 
 ,\:\:\Psi^{(k)}(t)\:\: }{\longrightarrow} & \left. \chi^{(k)}(T) \right] \\
                   & & & & & \left[ \Psi^{(k)}(0) \right. & \overset{{\boldsymbol \epsilon}^{(k)}(t)}
{\longrightarrow} & \left. \Psi^{(k)}(T) \right] \\
                   & & & & & \Psi^{(k+1)}(0) & \overset{{\boldsymbol \epsilon}^{(k+1)}(t)}
{\longrightarrow} & \Psi^{(k+1)}(T) .\\
\end{array}
\end{equation}
and calculate the laser field 
\begin{eqnarray}
\label{feld2}
\epsilon_j^{(k+1)}(t) &=& (1-\xi)\widetilde{\epsilon}_j^{(k)}(t)
- \frac{\xi}{\alpha_j}\im\, \langle\chi^{(k)}(t)|\hat{\mu}_j|\Psi^{(k+1)}(t)\rangle, \,\,\ j=x,y.\,\,\,\,\,\,\,\,\, 
\end{eqnarray}
If we want to avoid storing $\chi^{(0)}(t)$ in the memory we have to perform an additional forward propagation which in turn requires the knowledge of $\Psi^{(0)}(t)$. These extra propagations are indicated by the expressions in brackets. After the time evolution is complete we can close the loop and continue with equation \eref{eq:tdctrlstepk}.

The whole scheme can be depicted symbolically by:
\begin{equation}
\scriptsize
\label{eq:tdctrlscheme}
\begin{array}{l c c l c c c c }
  {\mbox{step 0:}} \,\, & \Psi^{(0)}(0) & \overset{{\bepsilon}^{(0)}(t)}{\longrightarrow} &
 \Psi^{(0)}(T) &  &  &  &\\
 {\mbox{step k:}} \,\, & & & \left[ \Psi^{(k)}(T) \right. & \overset{{\boldsymbol \epsilon}^{(k)}(t)}{\longrightarrow} &
 \left. \Psi^{(k)}(0) \right] & & \\
                   & & & \chi^{(k)}(T) & \overset{\widetilde{{\boldsymbol \epsilon}}^{(k)}(t) 
,\:\:\Psi^{(k)}(t)\:\: }{\longrightarrow} & \chi^{(k)}(0) & &\\
                   & & & & & \left[ \chi^{(k)}(0) \right. & \overset{\widetilde{{\boldsymbol \epsilon}}^{(k)}(t) 
 ,\:\:\Psi^{(k)}(t)\:\: }{\longrightarrow} & \left. \chi^{(k)}(T) \right] \\
                   & & & & & \left[ \Psi^{(k)}(0) \right. & \overset{{\boldsymbol \epsilon}^{(k)}(t)}
{\longrightarrow} & \left. \Psi^{(k)}(T) \right] \\
                   & & & & & \Psi^{(k+1)}(0) & \overset{{\boldsymbol \epsilon}^{(k+1)}(t)}
{\longrightarrow} & \Psi^{(k+1)}(T) .\\
\end{array}
\normalsize
\end{equation}
%
%
%
%
%
%
%
%
%
%
%
%
%
%

The main difference between this iteration and the schemes used in Ref.\ \cite{MT2003} is that one needs to know the time-dependent wave function $\Psi(t)$  to solve the inhomogeneous equation (\ref{eq:SOL_INHSE}) for the Lagrange multiplier $\chi(t)$. 

The choice of $\eta$ and $\xi$ completes the algorithm. $\xi = 1$ and $\eta = 1$ correspond to the algorithm suggested in Ref.\ \cite{ZR98}, while the choice $\xi = 1$ and $\eta = 0$ is analogous to the method used in Ref.\ \cite{K89} with a direct feedback of $\Psi^{(k)}(t)$. Further choices are discussed in Ref.\ \cite{MT2003}. A more detailed discussion on the convergence of this algorithm with exclusively time-dependent target operators and a modified version of the target functional can be found in Ref.\ \cite{ioanadipl}.

\subsection{Algorithms for time-dependent control targets with constraints}
\label{sec:tdtargetsconstraints}
%
%
For time-dependent targets a spectral restriction of the laser pulse turns out to be even more important than in the time-independent case. Besides the need to incorporate experimental limitations, spectral restrictions become important already at the level of modelling. For instance, assume that we want to optimize a time-dependent occupation or density such that it oscillates at a given frequency but the optimized pulse is not allowed to contain this frequency, e.g., to optimize high-harmonic generation, in that case a spectral constraint becomes inevitable. 

In the following we present an extension of the algorithm discussed in \sref{sec:ftheory} to incorporate time-dependent targets. During the presentation in \sref{sec:ftheory} we have distinguished between targets with a constraint on the fluence, filter algorithms using a penalty factor, and fluence constraints together with filter algorithms. Here we combine all of them in one scheme: 
\begin{equation}
\label{eq:tdctrlconstraints}
\begin{array}{l c c l c c c c }
  {\mbox{step k:}} \,\, & \Psi^{(k)}(0) & \overset{{\bepsilon}^{(k)}(t)}{\longrightarrow} &
 \Psi^{(k)}(T) &  &  &  &\\
          & & & \left[ \Psi^{(k)}(T) \right. & \overset{{\bepsilon}^{(k)}(t)}{\longrightarrow} &
 \left. \Psi^{(k)}(0) \right] & & \\
                   & & & \chi^{(k)}(T) & \overset{\widetilde{{\bepsilon}}^{(k)}(t),\Psi^{(k)}(t) 
}{\longrightarrow} & \chi^{(k)}(0), & &
\end{array}
\end{equation}
with the laser fields $\bepsilon^{(k)}(t),\tilde{\bepsilon}^{(k)}(t)$ given by
\begin{eqnarray}
\label{eq:feld1}
\widetilde{\epsilon}_j^{(k)}(t) &=& -  \frac{1}{\alpha_j^{(k)}}\im\langle\chi^{(k)}(t)|\hat{\mu}_j|\Psi^{(k)}(t)\rangle,\\
\bar{\epsilon}_j^{(k)}(t) &=&   \mathcal{G} \left[ \widetilde{\epsilon}_j^{(k)}(t)\right],  \\
\label{eq:feld2}
\epsilon_j^{(k+1)}(t) &=& \frac{\alpha_j^{(k)} }{\alpha_j^{(k+1)}} \bar{\epsilon}_j^{(k)}(t), \qquad \qquad j=x,y,
\end{eqnarray}
where $\mathcal{G}$ is a generic operator. Its precise form depends on the application, several examples can be found in \sref{sec:ffilter}. In the case of a fixed fluence the Lagrange multiplier $\alpha^{(k+1)}_j$ is defined by
\begin{eqnarray}
\label{eq:alpha1}
\alpha_j^{(k+1)} = \sqrt{\frac{\int_0^T \!\! dt  \left[ \alpha_j^{(k)} \bar{\epsilon}_j^{(k)}(t) \right]^2 }{E_{0_j}}},
\end{eqnarray}
otherwise (when $\alpha_j$ is a penalty factor) we set $\alpha^{(k+1)}_j = \alpha_j$.

The initial conditions in every iteration step are
\begin{eqnarray}
\nonumber
\label{eq:initial_psi}
  \Psi({\bf r},0)&=& \phi_i({\bf r}),\\
\label{eq:initial_chi}
  \chi({\bf r},T) &=& 0.
\end{eqnarray}
If only the fluence has to be kept fixed to a given value we use $\mathcal{G}[\epsilon(t)] = \epsilon(t)$.
%
%
\subsection{Example:  Optimal Control of time-dependent occupation numbers}
\label{sec:tdoct_example}
%
%
In this section we present an application of the algorithms (\ref{eq:tdctrlscheme}) and (\ref{eq:tdctrlconstraints}).  
We want to control the occupation numbers of the asymmetric double well model in time. We look for a laser pulse which drives the system as close as possible along a given path in quantum number space, defined as follows: 
First, excite the system from the ground state to the fourth excited state, then dump the occupation as fast as possible to the third excited state [realized by the step function $\Theta(5\,T/8 - t)$ in equation \eref{eq:c4}], and finally transfer the occupation to the first excited state.  In mathematical terms, the target occupation numbers $|c_n(t)|^2$ are defined as follows:
\bea
\label{eq:c0} 
c_0(t) &=&  \theta( T/2 -  t)\, \cos(\pi t/T)\\
\label{eq:c1} 
c_1(t) &=&  \theta(t - 3 T/4 )\, \sin(2 \pi t/T - 3 \pi /2)\\
\label{eq:c2} 
c_2(t) &=& 0\\ 
\label{eq:c3} 
c_3(t) &=&  \theta(t -  T/2)\, \left( 1- |c_0(t)|^2 - |c_1(t)|^2 - |c_4(t)|^2 \right)^{1/2}\\
\label{eq:c4} 
c_4(t) &=& \theta(T/2 -  t)\, \sin(\pi t/T) + \theta(t -  T/2)\, \theta(5\, T/8 -  t).
\eea
For $t < 0$ the system is in the ground state. The choice of the timings and the particular functional form of the population curves are just examples to demonstrate the capabilities of the optimization method.

\begin{figure}[!h]
\centering 
\subfigure[]{
  \label{fig:adwresultsfilter1}
  \resizebox{0.47\textwidth}{.26\textheight}{\includegraphics*{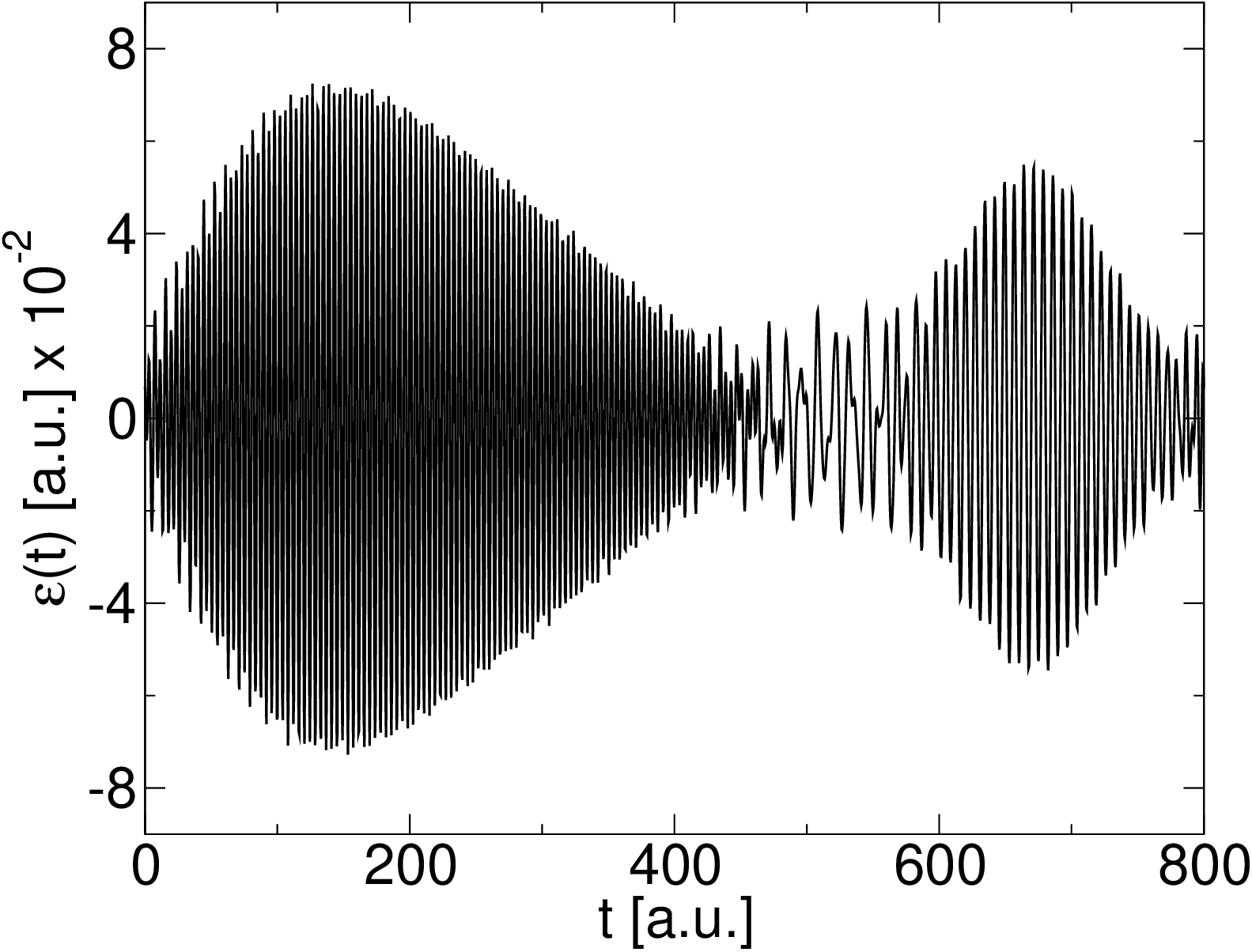}}
}
\subfigure[]{
  \label{fig:adwresultsfilter2}
  \resizebox{0.47\textwidth}{.26\textheight}{\includegraphics*{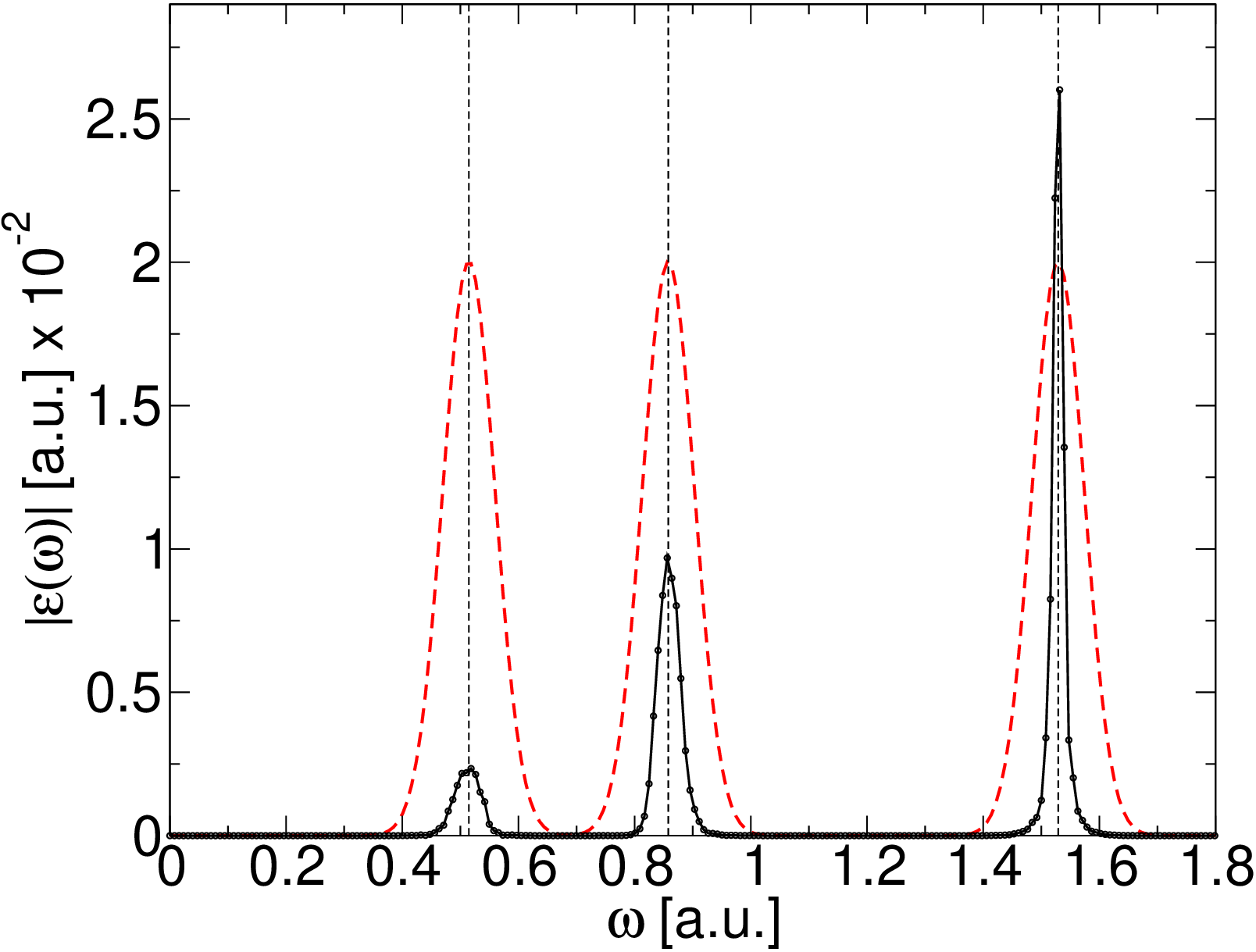}}
}
\\
\subfigure[]{
  \label{fig:adwresultsfilter3}
  \resizebox{0.47\textwidth}{.26\textheight}{\includegraphics*{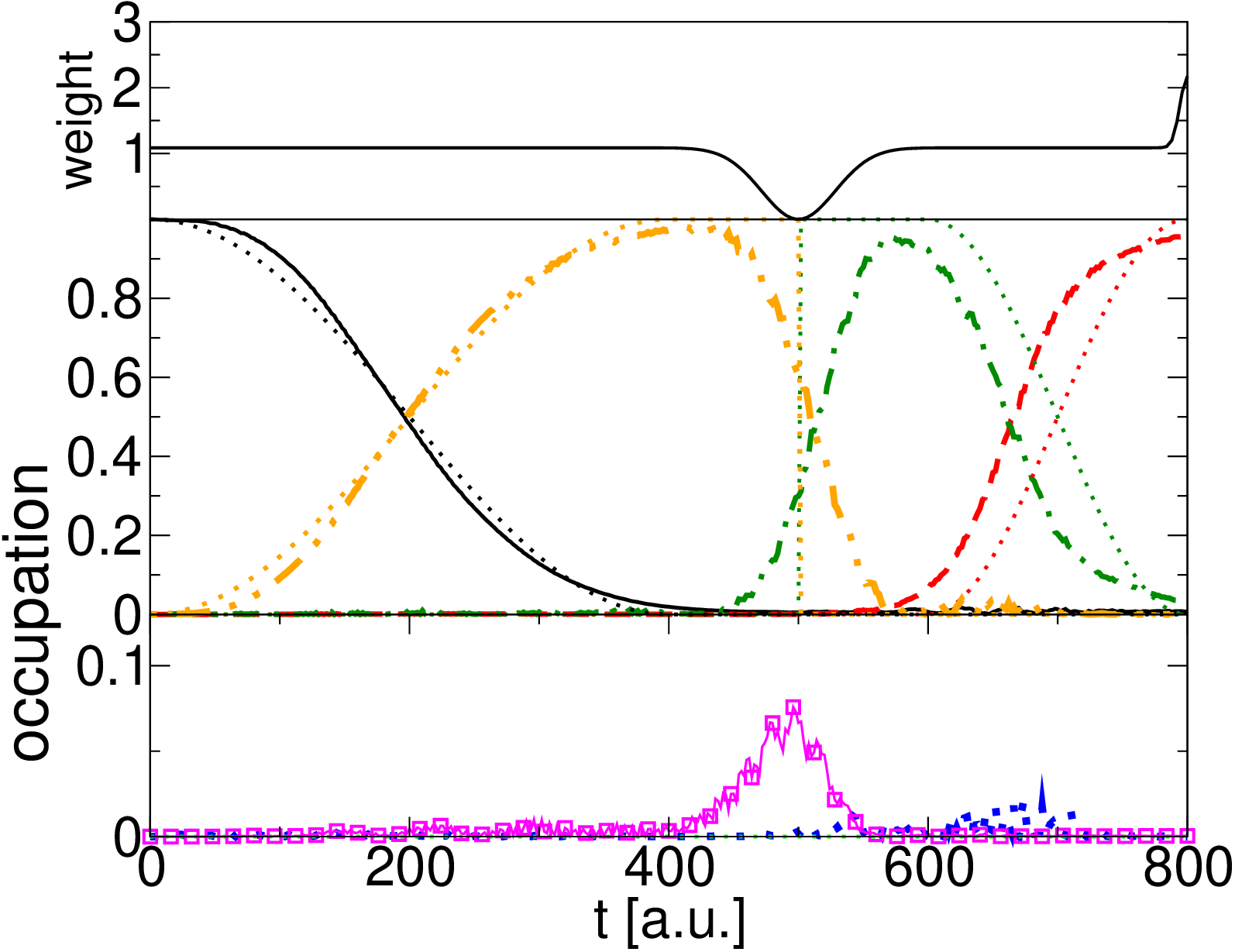}}
}
\subfigure[]{
  \label{fig:adwresultsfilter4}
  \resizebox{0.47\textwidth}{.26\textheight}{\includegraphics*{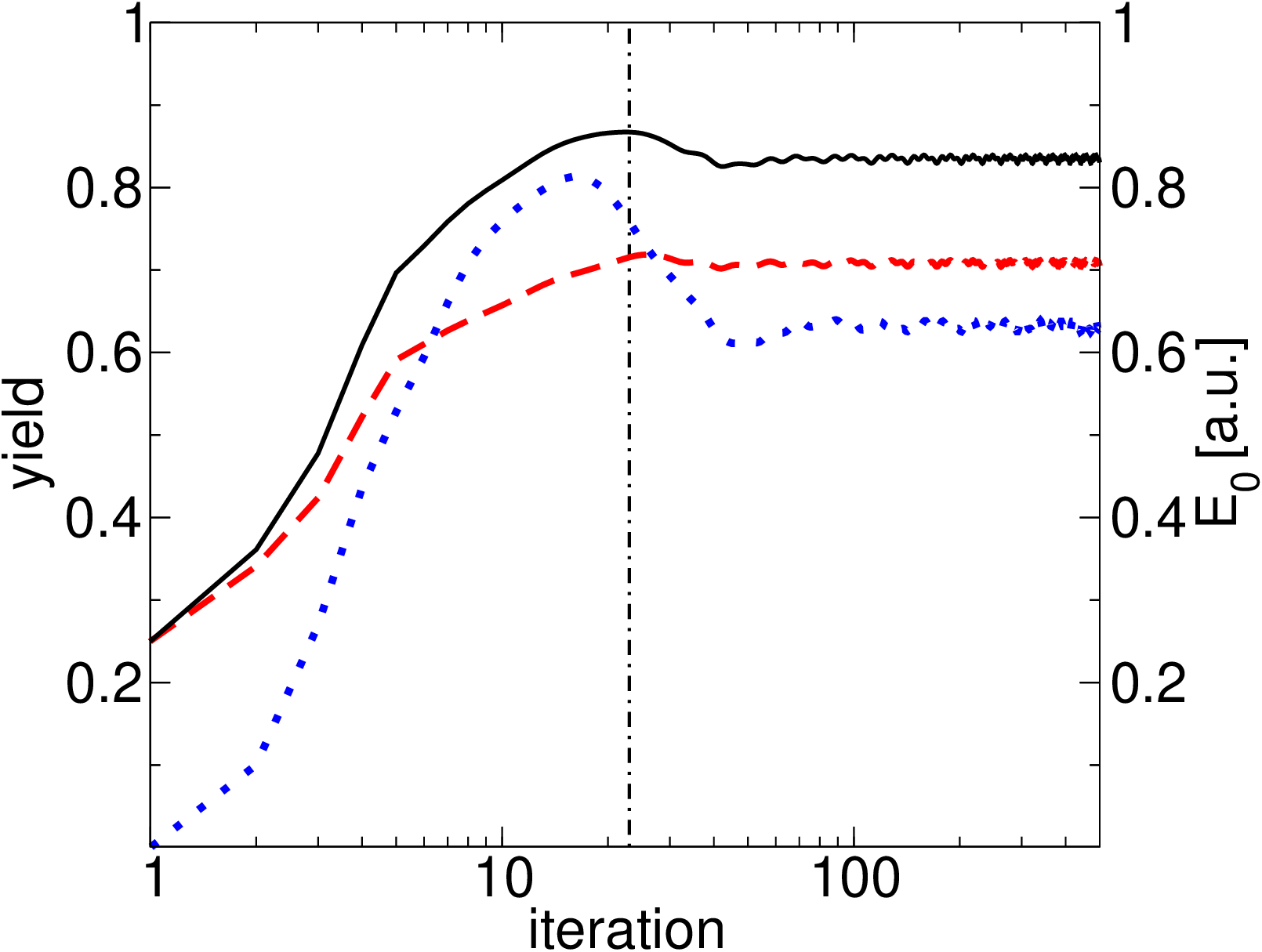}}
  }
\caption[Following a path in quantum number space with spectral constraints.]{(Color online). Control of time-dependent occupation numbers ($|0\ra\rightarrow |4\ra \rightarrow |3 \ra \rightarrow |1 \ra$). (a) Optimized laser field. (b) Spectrum of the optimized pulse together with the filter function [(\dashed); scaled by $0.02$]. (c) Top panel: Time-dependent weight function. Middle panel: Target and the achieved occupations $|\langle n | \Psi(t) \rangle |^2 $: $n=0$ (\full), $n= 1$ (\broken), $n=3$ (\chain), $n=4$ (\dashddot). The (\dotted) lines correspond to the target curves $|c_k(t)|^2$ from equations (\ref{eq:c0}-\ref{eq:c4}). Bottom panel:  $n=2$ (\dotted), $n=5$ (\opensquare). (d) Convergence of the algorithm, the (\full) line corresponds to the value of the target functional $J_1$, the (\dashed) line to $J$, and the (\dotted) line to the fluence of the pulse. The vertical line indicates the iteration with the maximum yield.}
\label{fig:adwresultsfilter}
\end{figure}

The squared moduli of the functions in equations (\ref{eq:c0})-(\ref{eq:c4}) are represented in the middle panel of \fref{fig:adwresultsfilter3} by the (\dotted) lines. The coefficients $c_n(t)$ define the target operator
\bea\nonumber
\hat{O} &=& | \phi(t) \rangle \langle \phi(t) |, \\
\nonumber
| \phi(t) \rangle &=& \sum_{n=0}^4 c_n(t) \rme^{-\rmi \mathcal{E}_n t } | n \rangle .
\eea
 The control objective has some resemblances with the one shown in \sref{sec:indirect031} where we have used a filter function to achieve a similar {\it path-selective excitation}. The time-dependent target operator, however, exhibits a much more precise control. We can specify not only the timings of the single transitions but also the rate. On the other hand, the choice of the trajectory does not require the same amount of intuition compared to the tracking method \cite{ZR2003}, i.e.\ with the help of OCT we can find a laser field that guides the system as close as possible along the trajectory, but not necessarily exactly. Therefore we are able to use also ``unphysical'' trajectories \cite{mySWG2005}, e.g., the step-like change at $t=5\,T/8$ in equation \eref{eq:c4}, where tracking methods will not work at all. 

In the following we present the results for the time-dependent target described above (see equations \eref{eq:c0} - \eref{eq:c4}). To suppress unwanted frequency contributions and to obtain a simple pulse shape we apply the filter algorithm described in \sref{sec:tdtargetsconstraints} with the parameters $\alpha = 0.2$, $w(t)=1$, $T=800$, $\epsilon^{(0)}(t)=0$, ($\gamma=1$, $\eta=1$).
The filter functions are Gaussian-shaped and centered around the field-free transition frequencies $\omega_{04}=1.5294$, $\omega_{43}=0.5147$, and $\omega_{31}=0.8580$. In addition, we want to improve the final occupation of the first excited state and to decrease the importance of the fast transition at $t=500$ by using a weight function which is shown in the top panel of \fref{fig:adwresultsfilter3} and defined by
\bea
\nonumber
f(t) &=& \left(1-\rme^{-\frac{(t- 5 T/8)^2}{1600}}+\rme^{-\frac{(t-T)^2}{64}} \right),\\
w(t) &=& T \frac{f(t)} {\int_0^T \!\! dt \, f(t)}.
\eea

Running the optimization the target functional $J_1$ reaches $86.72\%$ of its maximum value after $23$ iterations. The convergence is presented in \fref{fig:adwresultsfilter4}. It shows a fast convergence within the first ten iterations. Also shown in the inset of \fref{fig:adwresultsfilter2} is the increase of the fluence during the iterations. At the end of the iteration the laser fluence has reached $E_0 = 0.7586$.   

The optimized laser pulse is shown in \fref{fig:adwresultsfilter1} and its spectrum in \fref{fig:adwresultsfilter2}. Obviously, the spectrum shows only the three permitted frequency contributions $\omega_{04}$, $\omega_{43}$, and $\omega_{31}$ indicated by the vertical lines.

The system driven by the optimized laser follows the target population, indicated by the (\dotted) lines in \fref{fig:adwresultsfilter3}, very closely.

 Surprisingly, we observe an occupation of the $5$th excited state, indicated by the magenta (\opensquare) in \fref{fig:adwresultsfilter3}, although the resonance frequency $\omega_{35}$ is now suppressed by the filter. The excitation can only be understood by a mixing of two frequencies, namely $\omega_{04} - \omega_{43}= 1.0147 \approx \omega_{35}$. A time-frequency analysis shows that both are present at the time the occupation occurs. 

The weight function improves the occupation of the first excited state at the end of the pulse. The effect of the weight function on ``smoothing'' the transition at $t=500$ is rather small compared to runs without the weight function \cite{janphd}.


We conclude the analysis with the remark that for the optimizations of time-dependent targets we find the choice of the initial guess more important compared to the time-independent case. The simplest choice $\epsilon(t)=0$ turns out to be the best in this case, i.e., it produces the highest yields. Other choices end up in similar results but slightly smaller yields. The reason is a slowly decaying influence of the initial guess during the iteration. 
\section{Summary and Outlook}
\label{sec:conclusion}
%
In this tutorial we have presented the basics of quantum optimal control theory. We started with the derivation of the control equations from a suitably formed functional and then presented several algorithms to solve these equations. 
In addition we have presented a filter algorithm and the extension to time-dependent targets. Each section contained an example to demonstrate the discussed algorithm.

Currently, we are working on the implementation of several optimal control algorithms into the freely available software package \octopus~\cite{octopus,octopusurl}.  All algorithms in this article are already accessible within \octopus~for single-particle calculations in real-space.
In the future this implementation will allow us to investigate the optimal control of multi-electron systems using time-dependent density-functional theory (TDDFT)\cite{janphd}.


Besides further developments in experimental and theoretical capabilities it is extremely important for the future of quantum control to work on the interface between both ``worlds''. Namely to extract the most relevant information from the theoretical calculations and directly apply them in the experiment. 

\appendix
\section*{Appendix}
\setcounter{section}{1}
%
\subsection{Closed-loop learning experiments}
\label{sec:cllexperiments}

In the following we describe the principles of a closed-loop quantum control experiment using femtosecond laser pulses. In particular, we will focus on the process of laser pulse shaping which provides the connection with the theory presented here.
These quantum control experiments have become possible due to the improvement of laser pulse shaping \cite{WLPW92,BKPSG2005,SWYYH2005} and the implementation of closed-loop learning (CLL) techniques \cite{JR92}. Experiments using CLL have delivered highly encouraging results, ranging from the control of chemical reactions \cite{A98,LMR2001,D2003,B97,BDNG2001,HWCZM2002,BDKG2003,VKNNG2005} to the control of high-harmonic generation \cite{B2000,PWWSG2005}.


Consider a molecule which consists of three parts: $A-B-C$. The objective is to optimize a laser pulse that breaks the bond between $A$ and $B$, or in other words, maximizes the yield of $B-C$ fragments over $A-B$ fragments. 
The optimization proceeds iteratively within a learning loop. The loop starts with using an initial guess for the laser pulse. The pulse hits the molecules (in the gas phase) in a reaction chamber.  After the laser interaction the product is analyzed with a mass-spectrometer. 
 The resulting mass spectrum is then fed back into a computer which generates a new pulse shape. The prediction of the new pulse shape is usually done by a genetic algorithm \cite{ZFKM2001}. When the reaction chamber is loaded with a new sample of the molecule the procedure starts again. The loop is continued until the best pulse is found. 

To understand some of the problems in these control experiments we have to take a closer look at the process of pulse shaping. The first point to realize is that a femtosecond pulse is broad in frequency space, i.e., many frequency components are needed to form this pulse ($20~\mathrm{fs} \sim 1000~\mathrm{cm}^{-1}$). The idea of pulse shaping is to manipulate the phases and the amplitudes of these frequencies which can be achieved in the following way: The incoming pulse is targeted on a grating which separates the frequencies in space. Then the light beam enters a liquid crystal modulator (LCM or SLM: spatial light modulator) which consists of several (typically 128 to 640) small ``windows'' (pixels). Every pixel modifies the amplitude and phase of the incoming light separately, controlled by the computer algorithm. The transmitted beam  is then transformed back using a second grating. The first loop of the experiment has to be started with an initial guess for the settings of the LCM. Since the convergence of the genetic algorithm (speed and final result) might depend on the initial guess a good choice is very important since the search space itself is gigantic:
Let us assume a resolution of the amplitude and the phase in each LCM pixel of 4 bit which corresponds to $2^{4} = 16$ different settings. With $128$ Pixels we have $128^{(2 \cdot 16)}\approx 10^{67}$ different pulses that can be generated. If we fix the amplitude setting and use {\it phase-only shaping}, we can reduce the number of configurations by a factor $1/2$. If we give up the idea of free optimization and assume a function which describes the settings of the LCM with $n$ parameters \cite{ZFKM2001} we can reduce the search space from $128$ to $n$ dimensions and therefore $n^{(2 \cdot 16)}$ configurations. 
To determine a good parameterization of the pulse, theoretical calculations which take into account the experimental constraints are extremely important.
%

%
%
\subsection{Two-Level theory}
\label{sec:TLStheory}
%
Two-level systems are applied in many fields of physics from spin models to the area of quantum computation \cite{NC2004}. Since they are analytically solvable  if the rotating-wave approximation (RWA) is used, they often serve as simple but powerful models to analyze field-matter interaction problems. The availability of exact solutions is also our motivation to study the system. The idea is to illustrate the OCT algorithm and to compare the results to the exact solutions within the RWA. 

First an introduction to the theoretical concepts of two-level systems is given. We start by reviewing the RWA and then we derive two recipes for population inversion within the RWA ($\pi$-pulses) and in first-order perturbation theory. 
We conclude this brief review with a controllability analysis in \ref{sec:cc2level}. 

\subsection{Two-level system: Basics}
\label{sec:2lev}

Let our model system consist of two orthonormal states $|a \ra$ and $|b \ra$. The state vector at any time $t$ is given by
\bea
\label{eq:statevec}
|\Psi(t) \ra = c_a(t)|a \ra   + c_b(t)|b \ra =: 
\left(
\begin{array}{c} 
  c_a(t) \cr
  c_b(t) \cr
\end{array}
\right) .
\eea
The time evolution of $|\Psi(t)\ra$ is described by the \TDSE~
\bea
\nonumber
\rmi \partial_t |\Psi(t) \ra = \hat{H}(t) |\Psi(t) \ra,
\eea
with the Hamiltonian in the basis $|a \ra$ and $|b \ra$ given by
\bea
\label{eq:2lev_system}
\hat{H}(t) = \left(
\begin{array}{cc}
  \omega_a & 0 \cr 
    0              &  \omega_b 
\end{array}
\right) - \epsilon(t)
\left( 
\begin{array}{cc}
\rho_{aa}  & \rho_{ab} \cr                
\rho_{ba}  &  \rho_{bb} 
\end{array}
\right) ,
\eea
where we assume $\rho_{ab} = \rho_{ba} = \mu$ and $\rho_{aa} = \rho_{bb}=0$. The \TDSE~then yields the following system of differential equations,
\bea
\label{eq:ode_ca_general}
\dot{c}_a &=& -\rmi \omega_a c_a(t) + \rmi \epsilon(t) \mu c_b(t),\\
\label{eq:ode_cb_general}
\dot{c}_b &=& -\rmi \omega_b c_b(t) + \rmi \epsilon(t) \mu c_a(t).
\eea
Applying the transformation $g_k(t)=\exp(\rmi \omega_k t) c_k(t)$ we obtain
\bea \label{eq:ode_ga_general}
\dot{g}_a &=&  \rmi \epsilon(t) \mu \rme^{- \rmi \omega_{ba} t} g_b(t), \\
\label{eq:ode_gb_general}
\dot{g}_b &=&  \rmi \epsilon(t) \mu \rme^{  \rmi \omega_{ba} t} g_a(t),
\eea 
where $\omega_{ba} =  \omega_b  - \omega_a$.

For an arbitrary laser field $\epsilon(t)$, the set of differential equations (\ref{eq:ode_ga_general}) and  (\ref{eq:ode_gb_general}) is only numerically solvable. 

\subsubsection{Rotating wave approximation (RWA)}
%
If the laser field $\epsilon(t)$ is chosen to be 
\bea \label{eq:field_exp}
\epsilon(t) = A \sin( \nu t ) = \frac{A}{2 \rmi} \big( \rme^{\rmi \nu t} - \rme^{-\rmi \nu t} \big),
\eea
we can rewrite the differential equations \eref{eq:ode_ga_general} and \eref{eq:ode_gb_general} as
\bea
\nonumber
\dot{g}_a(t)& =& \rmi \Omega_R  \, \rme^{ - \rmi \omega_{ba} t} \frac{1}{2 \rmi} \left(  \rme^{\rmi \nu t} - \rme^{-\rmi \nu t} \right) g_b(t)\\
\label{eq:ode_ga_special}
& =& \frac{\Omega_R}{2} \big(  \rme^{ - \rmi (\omega_{ba} - \nu) t} -  \rme^{- \rmi (\omega_{ba} + \nu)  t} \big) g_b(t) ,
\eea
where we have introduced the expression $\Omega_R = A \mu$ known as the {\it Rabi-frequency}. 
Analogously, we get
\bea
\label{eq:ode_gb_special}
\dot{g}_b(t)  =   \frac{- \Omega_R}{2} \big( \rme^{-\rmi  \Delta_{-} t} - \rme^{-\rmi \Delta_{+} t} \big) g_a(t) 
\eea
with $ \Delta_{\pm} = (\omega_{ba} \pm \nu) $.
If $\omega_{ba}$ and $\nu$ are of similar magnitude (and have the same sign), i.e., $\Delta_{-}$ is small, we may neglect the term
\bea
 \rme^{-\rmi \Delta_{+} t}=\rme^{ - \rmi (\omega_{ba} + \nu)  t},
\eea
 because it oscillates much faster than the other term and cancels out on average. This approximation is called the rotating wave approximation (RWA).

\subsubsection{Solutions: Resonant case in RWA}
%
Using the RWA, equations \eref{eq:ode_ga_special} and \eref{eq:ode_gb_special} can be solved analytically \cite{SZ97}. 
The solution is
\bea
\nonumber
g_a(t) &=& \left[ g_a(0) \left( \cos(\Omega t /2) - \frac{\rmi \Delta_{-}}{\Omega} \sin (\Omega t /2) \right)  \right. \\ 
\label{eq:full_solution_ga} 
&& \qquad \left. - \frac{\Omega_R}{\Omega} g_b(0) \sin (\Omega t /2) \right] \rme^{\rmi \Delta_{-} t /2} , \\
\nonumber
g_b(t) &=& \left[ g_b(0) \left( \cos(\Omega t /2) + \frac{\rmi \Delta_{-}}{\Omega} \sin (\Omega t /2) \right) \right. \\ 
\label{eq:full_solution_gb} 
&& \qquad \left. - \frac{\Omega_R}{\Omega} g_a(0) \sin (\Omega t /2) \right] \rme^{-\rmi \Delta_{-} t /2} ,
\eea
with
\bea
\nonumber
\Omega^2 &=& \Omega_R^2 + \Delta_{-}^2 .
\eea
For the initial state, we choose $g_a(0) = 1$ and $g_b(0)=0$, i.e., the population is assumed to rest in the lower-state at $t=0$. Thus, the occupation numbers are given by
\bea
\label{eq:occRWAa}
|g_a(t)|^2 &=& \cos^2(\Omega \,t /2) + \frac{\Delta_{-}^2}{\Omega^2} \sin^2(\Omega \, t /2), \\
\label{eq:occRWAb}
|g_b(t)|^2 &=& \frac{\Omega_R^2}{\Omega^2} \sin^2(\Omega \, t /2).
\eea
The populations oscillate with the frequency $\Omega /2$ which, in the resonant case $\Delta_{-}=0$, corresponds to half the Rabi frequency $\Omega_R/2$. On the other hand, if we fix the propagation time to $T$ and look at the final populations for different field amplitudes, we also find oscillations of $|g_a(T)|^2$ and $|g_b(T)|^2$ versus the field amplitude. These are called Rabi oscillations as well. 

\subsubsection{RWA: Optimal amplitudes}
%
If we want to maximize the occupation of state $|b\rangle$ at the final time $T$, it is possible to find an optimal amplitude $A$ by looking at the structure of equation \eref{eq:occRWAb}, i.e., if $|g_b(T)|^2$ reaches its maximum at $T$ it has to be zero at $2T$: $|g_b(2T)|^2 = 0$.
%
%
 %
Therefore, we have
\bea
\nonumber
 && \sin(\Omega\, T) = 0, \\
\nonumber
 \Rightarrow && \Omega T = \pi (2k +1) \,\,\, \mbox{with} \,\,\, k=0,1,\ldots, \\
\label{eq:optampeq}
 \Rightarrow && \Omega_R^2  = \frac{(2k+1)^2 \pi^2}{T^2} - \Delta_{-}^2, 
\eea
where $k$ tells us how many times the maximum occupation has been reached within the time interval $[0,T]$.
From equation \eref{eq:optampeq} follows the optimal amplitude
\bea
\label{eq:opt_amp_rwa_off}
A = \frac{1}{\mu} \sqrt{\frac{(2k +1)^2 \pi^2}{T^2} - \Delta_{-}^2}.
\eea
Note that for a fixed final time $T$ the number $k$ has to be chosen large enough, so that the root does not yield an imaginary number for the value of $\Delta_{-}$.
The occupation of state $| b \rangle$ for the optimal amplitude in equation \eref{eq:opt_amp_rwa_off} is 
\bea
\max |g_b(T)|^2 = \frac{\Omega_R^2}{\Omega^2} = 1- \frac{\Delta_{-}^2 T^2}{(2k+1)^2 \pi^2}.
\eea 
For a fixed final time $T$, the result will improve with larger $k$ and therefore with increasing field strength [see equation \eref{eq:opt_amp_rwa_off}]. 

The solution in the resonant case ($\Delta_{-}=0$),
\bea
\label{eq:opt_amp_res}
 \stackrel{k=0}{\Rightarrow} & A_{\mathrm{RWA}}  &= \frac{\pi}{T \mu }\, ,
\eea
plays an important role in this work since it will be used as a benchmark for optimal control solutions. 

The results, equations \eref{eq:opt_amp_rwa_off} and \eref{eq:opt_amp_res}, allow us to find a pulse which maximizes the occupation in the upper state $|b \rangle$ in an arbitrarily short time. In the limit $T \to 0$ the optimal amplitude diverges as $1/T$. However, if $T$ becomes comparable to $1/\nu$ the RWA is no longer applicable. Besides this limitation, one has to take care that the full quantum system can be approximated by a two-level model at all in the case of the resulting strong fields, i.e., usually more levels have to be considered.
%

\subsubsection{Results from perturbation theory}
\label{sec:perturb2lev}
In the following we determine an optimal laser field for population inversion with the help of perturbation theory. We consider the two-level system defined by equations \eref{eq:ode_ca_general} and \eref{eq:ode_cb_general}. In time-dependent perturbation theory \cite{BRANSDEN} the $k$th-order coefficients are determined by
\bea
\label{eq:ode_cb_kth}
\dot{c}_b^{(k)} &=&  - \rmi  \mu \epsilon(t) \rme^{\rmi \omega_{ba} t } c_a^{(k-1)}, \\
\label{eq:ode_ca_kth}
\dot{c}_a^{(k)} &=&  - \rmi  \mu \epsilon(t) \rme^{-\rmi \omega_{ba} t } c_b^{(k-1)}.
\eea
Now consider a functional $J = J_1 + J_2$ with
\bea
J_1 &=& | \la b | \Psi(T) \ra |^2, \\
J_2 &=& - \lambda \int_0^T \!\! dt \,\, \epsilon^2(t) ,
\eea
where $\lambda$ is a given penalty factor.\\
This functional expresses our wish to transfer the occupation, initially in state $|a\ra$, to state $|b\ra$ at the end of the laser pulse and to minimize the fluence of the laser.

\begin{figure}[!h]
\centering 
\subfigure[]{
  \label{fig:2lev_pertA}
  \resizebox{0.48\textwidth}{.26\textheight}{\includegraphics{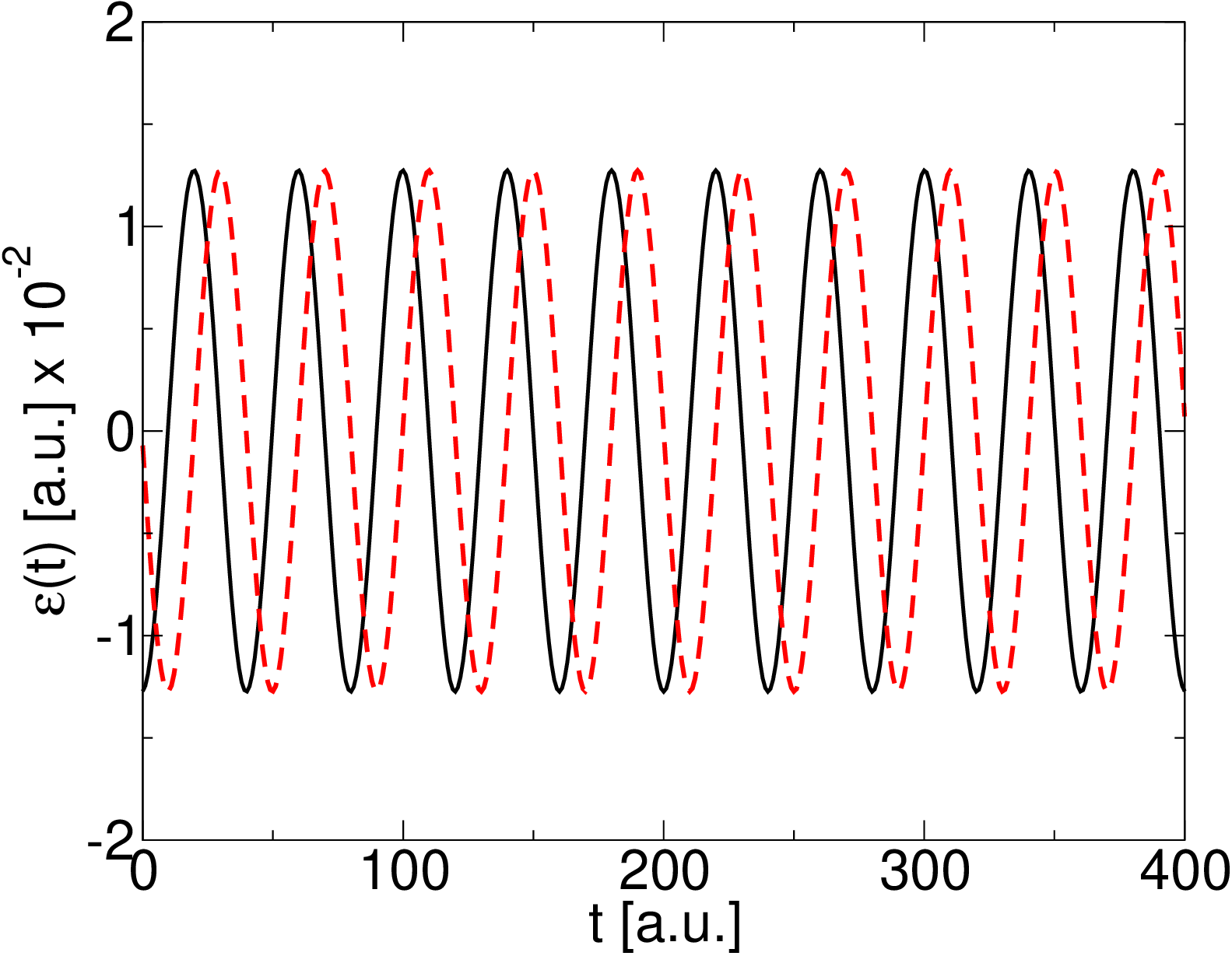}
    }}
\subfigure[]{
  \label{fig:2lev_pertB}
  \resizebox{0.48\textwidth}{.26\textheight}{\includegraphics{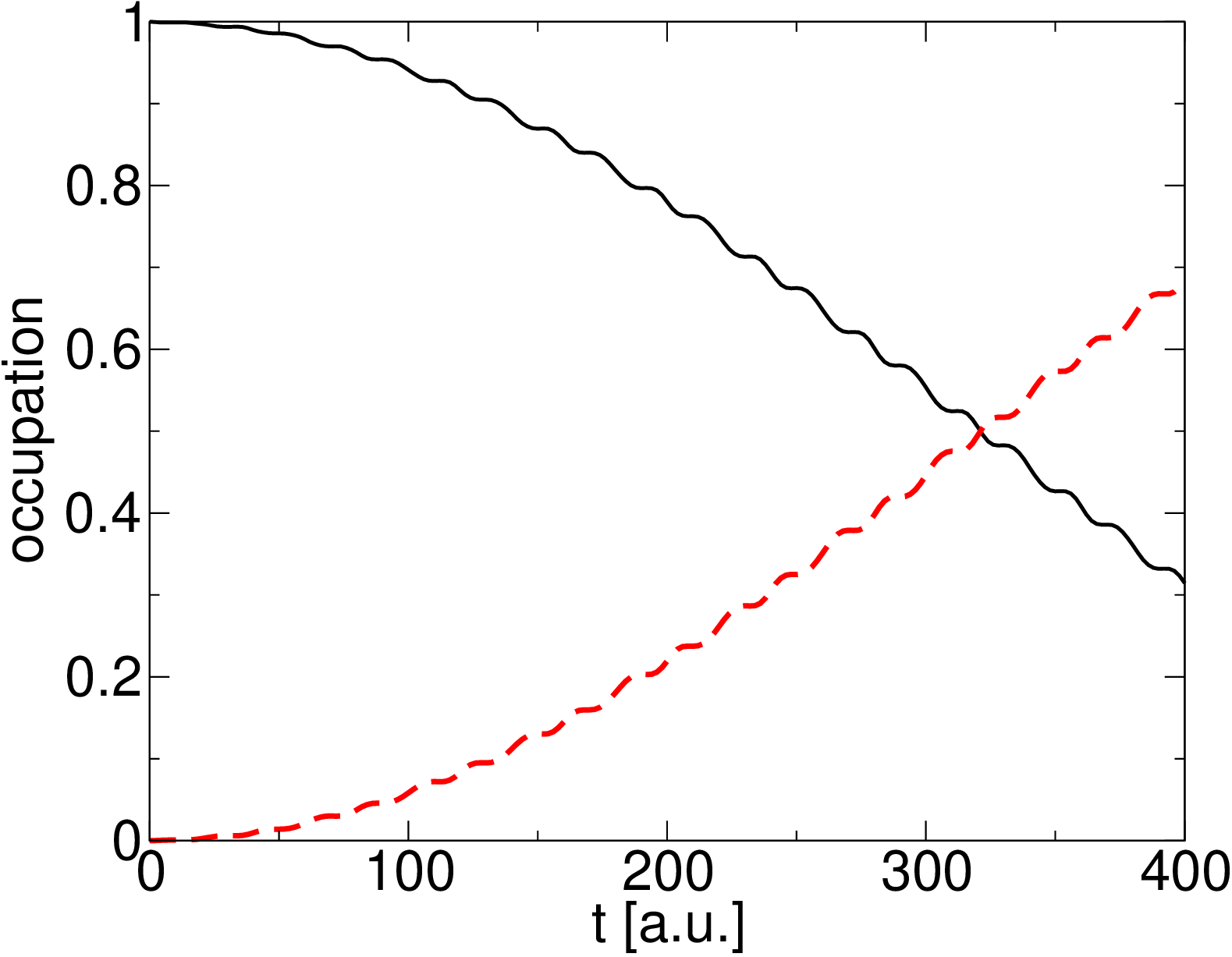}}
    }
\caption[Results from perturbation theory.]{(Color online). (a) The two largest eigenvectors of equation \eref{eq:perturbEV}. The (\full)~line corresponds to an eigenvalue of $30.8478$ while the (\dashed)~line corresponds to the eigenvalue $30.8032$. The third largest eigenvalue is $1.1126\,\cdot 10^{-14}$. The shown eigenvectors are oscillations with $\omega=\omega_{ba}=0.1568$ and have an amplitude of $A=0.1273$, after rescaling it with equation \eref{eq:rescale}. 
If we propagate the two-level system with the field corresponding to the largest eigenvalue we find $J_1^{(1)} = 0.6850$ [see (b); (\dashed) line], showing that the first-order approximation is insufficient in this case. The (\full) line corresponds to the ground state occupation.} 
\label{fig:2lev_pert}
\end{figure}

Setting the variation of this functional to zero yields
\bea
\label{eq:oct_eq}
\frac{\delta J_1 }{\delta \epsilon(t)} = -2 \lambda \epsilon(t).
\eea
Now we express $J_1$ in first-order perturbation theory using equations \eref{eq:ode_cb_kth},
\bea
\nonumber
 \la b | \Psi(T) \ra^{(1)} = c_b^{(1)}(T) = \mu  \int_0^T \!\! dt' \,  \rme^{\rmi \omega_{ba} t'} \epsilon(t') ,
\eea
and obtain
\bea
 J_1^{(1)} = \half |\mu|^2  \int_0^T \!\! dt' \!\! \int_0^T \!\! dt'' \, \cos\left[\omega_{ba} (t'-t'')\right] \epsilon(t') \epsilon(t'') .
\eea
The functional derivative of $J_1^{(1)}$ with respect to the laser field is
\bea
\frac {\delta J_1^{(1)}} {\delta \epsilon(t)} =  |\mu|^2 \int_0^T \!\! dt' \, \cos\left[\omega_{ba} (t'- t)\right] \epsilon(t') ,
\eea
and the optimal field can be determined by 
\bea
\label{eq:perturbEV}
\lambda \epsilon(t) =\int_0^T \!\! dt'  |\mu|^2 \cos\left[\omega_{ba} (t'- t)\right] \epsilon(t').
\eea
This equation represents an eigenvalue problem which also means that the penalty factor $\lambda$ does only yield solutions for certain values, i.e., the eigenvalues of this equation. It can be interpreted as the yield $J_1^{(1)}$ per unit of field fluence \cite{KWWYM93,KMWY95}. Thus, the eigenvector corresponding to the largest eigenvalue yields the optimal field (see \fref{fig:2lev_pert}). 
%
%
However, the amplitude of that field is arbitrary. It can be fixed by the maximum value that we can achieve for the occupation of state $|b\ra$, which is one. Therefore, the optimal field is given by:
\bea
\label{eq:rescale}
\epsilon_{\mathrm{opt}}(t) =  \sqrt{\frac{1/\lambda}{\int_0^T \!\! dt \, \epsilon^2(t) }} \, \epsilon(t).
 \eea
An optimal field from the two-level system obtained in this way is shown in \fref{fig:2lev_pert}.
We emphasize that the optimal field obtained in this way is {\it unique} and the RWA need not be applied. However, as the example calculation shows, the amplitude obtained in this way is too small. The final occupation of the target state is $68.50\%$.

\subsection{Example: Complete controllability of two-level-system}
\label{sec:cc2level}
%
To show that the two-level system is completely controllable we use the theorems 
of Refs.\ \cite{RSDRP95,SFS2001}, which have been stated in \sref{sec:completecontrol}. Following these theorems, we have to construct the Lie algebra $L_0$ of the skew-Hermitian operators $\rmi H_0$ and $\rmi H_1$ describing our system. In particular, we have to show that the dimension of $L_0$ is $N^2=2^2$.
For the construction we will apply the approach of Ref.\ \cite{SFS2001}.
The calculation can be simplified by rewriting the Hamiltonian of the two-level system (\ref{eq:2lev_system}) as
\bea
\hat{H} = 
\frac{1}{2}  \left( \omega_a + \omega_b \right) \hat{I} + \frac{1}{2} \left( \omega_a - \omega_b \right) \sigma_z - \epsilon(t) \mu \sigma_x ,
\eea
where we have used $\mu = \rho_{ab} = \rho_{ba}$ and $\rho_{aa} = \rho_{bb} = 0$.
The $\sigma_k$ are the Pauli matrices which obey the commutation relation: $\left[\sigma_i, \sigma_j \right] = 2 \rmi \epsilon_{ijk} \sigma_k$. 
At this point we introduce the matrix $W$ which will represent the basis of $L_0$. If its rank is $N^2$ the system is completely controllable. The matrix consists of columns $W_{:,k}$, i.e.,
 $W=\left(W_{:,1}, W_{:,2}, \ldots , W_{:,N^2}\right)$. The columns are calculated by transforming the $N \times N$ matrix $H_k$ into a vector with $N^2$ components by appending the next column to the end of the previous one. The columns are calculated in the following way:
\begin{enumerate}
\item Set $W_{:,1} := \hat{H}_0$ to the unperturbed Hamiltonian, here $W_{:,1} = (\omega_a,0,0,\omega_b)^\dagger$.
\item Set $W_{:,j} := \hat{H}_j$ to the perturbation/control Hamiltonians, here $W_{:,2} = (0, \epsilon \mu, \epsilon \mu,0)^\dagger$.
\item Calculate all non-vanishing commutators of $\hat{H}_0$ and $\hat{H}_j$ and add to $W$ if it is linearly independent from the existing columns. Here, we have simply $W_{:,3} = (0,\omega_{ba} \epsilon \mu,-\omega_{ba} \epsilon \mu,0)^\dagger$.
\item After appending the new column, all commutators with the previous columns have to be evaluated and added (if linearly independent). \\
This will result in $W_{:,4} = (2 \epsilon^2 \mu^2 \omega_{ba} ,0,0,-2 \epsilon^2 \mu^2 \omega_{ba})^\dagger$.
\item Repeat steps 3 and 4 until the matrix is full, or no new linearly independent columns can be found, and then determine the rank of $W$.
\end{enumerate}
The two-level system results in the matrix 
\bea
\nonumber
W =
\left( 
\begin{array}{cccc}
\omega_a & 0            & 0 & 2\epsilon^2 \mu^2 \omega_{ba} \cr
      0  & \epsilon \mu & \omega_{ba} \epsilon \mu & 0 \cr
      0  & \epsilon \mu & -\omega_{ba} \epsilon \mu & 0 \cr
\omega_b & 0            & 0 & -2\epsilon^2 \mu^2 \omega_{ba} \cr
\end{array}
\right).
\eea
If $\omega_{b} \neq -\omega_{a}$ the rank of $W$ is $N^2=4$ and therefore the system is completely controllable. If $\omega_{b} = -\omega_{a}$ (i.e.\ $\omega_{ba} = -2\omega_a$) the rank reduces to $N^2 - 1 = 3$ and the system is {\it not} completely controllable according to the definition. 
In other words, the difference of being controllable or not depends on whether $\hat{H}_0$ has a non-zero trace or not. The disturbing fact is that a traceless $\hat{H}_0$ can be changed to a Hamiltonian with non-zero trace by shifting the energy levels. However, this shift is not of physical significance. The dilemma is resolved  by realizing that in the case of a traceless $\hat{H}_0$ we (only) loose control over the phase of the state, which is not relevant in the case of population inversion \cite{SFS2001}. 

%

\ack
We thank Elham Khosravi for reviewing the manuscript.
 This work was supported, in part, by the Deutsche Forschungsgemeinschaft through the SFB450, and the NANOQUANTA Network of Excellence of the European Union.

\section*{References}
\bibliographystyle{phpf}
\bibliography{mylib}
\end{document}